\documentclass[twocolumn]{aastex631}

\let\tablenum\relax
\usepackage{siunitx}
\DeclareSIUnit\year{yr}
\usepackage[version=4]{mhchem}
\usepackage{booktabs}
\usepackage{multirow}
\usepackage{tabularx}
\usepackage{array}

\accepted{November 10, 2022}

\acceptjournal{\textit{The Astrophysical Journal}}
\reportnum{Supporting figure sets are available: \href{https://doi.org/10.6084/m9.figshare.21545148}{figshare}, doi: \href{https://doi.org/10.6084/m9.figshare.21545148}{10.6084/m9.figshare.21545148}}

\shorttitle{Nucleobases and Volatiles in Meteorites}
\shortauthors{Paschek et al.}

\begin{document}

\title{Meteorites and the RNA World: Synthesis of Nucleobases in Carbonaceous Planetesimals and the Role of Initial Volatile Content}

\correspondingauthor{Klaus Paschek}
\email{paschek@mpia.de}

\author[0000-0003-2603-4236]{Klaus Paschek}
\affiliation{Max Planck Institute for Astronomy, K{\"o}nigstuhl 17, D-69117 Heidelberg, Germany}

\author[0000-0002-3913-7114]{Dmitry A. Semenov}
\affiliation{Max Planck Institute for Astronomy, K{\"o}nigstuhl 17, D-69117 Heidelberg, Germany}
\affiliation{Department of Chemistry, Ludwig Maximilian University of Munich, Butenandtstr.~5-13, House F, D-81377 Munich, Germany}

\author[0000-0002-5449-4195]{Ben K. D. Pearce}
\affiliation{Origins Institute and Department of Physics and Astronomy, McMaster University, ABB 241, 1280 Main Street, Hamilton, ON L8S 4M1, Canada}
\affiliation{Department of Earth and Planetary Science, Johns Hopkins University, Baltimore, MD 21218, USA}

\author[0000-0002-2665-3777]{Kevin Lange}
\affiliation{Institute for Theoretical Astrophysics and Center for Astronomy, Ruprecht Karl University of Heidelberg, Albert-Ueberle-Str.~2, D-69120 Heidelberg, Germany}
\affiliation{Anton Pannekoek Institute for Astronomy, University of Amsterdam, Science Park 904, 1098 XH Amsterdam, The Netherlands}

\author[0000-0002-1493-300X]{Thomas K. Henning}
\affiliation{Max Planck Institute for Astronomy, K{\"o}nigstuhl 17, D-69117 Heidelberg, Germany}

\author[0000-0002-7605-2961]{Ralph E. Pudritz}
\affiliation{Origins Institute and Department of Physics and Astronomy, McMaster University, ABB 241, 1280 Main Street, Hamilton, ON L8S 4M1, Canada}

\begin{abstract}\noindent
Prebiotic molecules, fundamental building blocks for the origin of life, have been found in carbonaceous chondrites. The exogenous delivery of these organic molecules onto the Hadean Earth could have sparked the polymerization of the first RNA molecules in Darwinian ponds during wet-dry cycles. Here, we investigate the formation of the RNA and DNA nucleobases adenine, uracil, cytosine, guanine, and thymine inside parent body planetesimals of carbonaceous chondrites. An up-to-date thermochemical equilibrium model coupled with a 1D thermodynamic planetesimal model is used to calculate the nucleobase concentrations. Different from the previous study \citep{Pearce2016}, we assume initial volatile concentrations more appropriate for the formation zone of carbonaceous chondrite parent bodies. This represents more accurately cosmochemical findings that these bodies have formed inside the inner, ${\sim \text{\SIrange{2}{5}{au}}}$, warm region of the solar system. Due to these improvements, our model represents the concentrations of adenine and guanine measured in carbonaceous chondrites. Our model did not reproduce \textit{per se} the measurements of uracil, cytosine, and thymine in these meteorites. This can be explained by transformation reactions between nucleobases and potential decomposition of thymine. The synthesis of prebiotic organic matter in carbonaceous asteroids could be well explained by a combination of i) radiogenic heating, ii) aqueous chemistry involving a few key processes at a specific range of radii inside planetesimals where water can exist in the liquid phase, and iii) a reduced initial volatile content (\ce{H2}, \ce{CO}, \ce{HCN}, \ce{CH2O}) of the protoplanetary disk material in the parent body region compared to the outer region of comets.
\end{abstract}


\section{Introduction}\label{sec:intro}

The origin of biomolecules on the early Earth is still a major question to those researching the emergence of life. RNA molecules are favored for being the crucial intermediate step toward the emergence of living systems \citep{Rich1962,Gilbert1986}. RNA is able to store genetic information encoded in the order of its monomers, the ribonucleotides, as well as to catalyze its own polymerization, and to self-replicate \citep[see, e.g.,][]{Kruger1982,Guerrier-Takada1983,Guerrier-Takada1984,Zaug1986,Cech1986,Johnston2001,Vaidya2012,Attwater2018,Cojocaru2021,Kristoffersen2022}.

There are generally two sites considered for the emergence of RNA: freshwater ponds and hydrothermal vents. How the abiotic polymerization of the first long and complex biomolecules could have occurred in the pristine oceans in the vicinity of hydrothermal vents is an open question, as necessary concentrations are unlikely to be reached \citep{Pearce2017}. Concentrated nitrogen (\ce{N2}) necessary for the nucleobase synthesis is probably missing, as most of it might be reduced to ammonium (\ce{NH4+}) due to the presence of hydrogen sulfide (\ce{H2S}) in this setting \citep{Schoonen2001}. As only little \ce{N2} is present, it is questioned if nucleobases could be formed \textit{in situ} in hydrothermal vents. Therefore, for the emergence of the first RNA world in hydrothermal vents, a delivery by an exogenous source would be necessary. Furthermore, cells today have different ion concentrations than the ocean water \citep[cells contain more \ce{K^+}, \ce{Zn^{2+}}, phosphate, and transition metals (e.g., \ce{Mn^{2+}}), and less \ce{Na^+};][]{Mulkidjanian2012}. As the first proto-cells likely did not possess ion-tight membranes nor membrane pumps, it is proposed that modern cells might reflect the ion concentration of proto-cells, and hence, of present-day or reconstructed primeval oceans. This is used as an argument to cast doubt upon whether hydrothermal vents at the bottom of the oceans were a likely place for the origin of life \citep[see, e.g.,][]{Deamer2019}.

It might be possible that RNA was not the first biomolecule that played the role of genetic material. Several other candidates have been proposed and studied, e.g., glycol nucleic acids \citep{Seita1972,Zhang2005}, polyamide nucleic acids \citep{Nielsen1991}, threose nucleic acids \citep{Schoening2000,Yu2012}, or different nucleobases \citep{Crick1968,Cafferty2015,Cafferty2016}. It is possible that chemical evolution started with the emergence of other types of information-coding molecules, which then eventually transitions into (proto-)RNA via an evolutionary process in the so-called pre-RNA world \citep[see, e.g.,][]{Hud2018}.

The idea to locate the origin of life in small freshwater basins or ponds on the first continental crust units or volcanic islands attracts more and more attention nowadays. Charles Darwin originally posited in 1871 that life might have emerged in a warm little pond (WLP) filled with simple prebiotic ingredients \citep{Deamer2017,doi:10.1089/ast.2019.2045}.
\citet{Pearce2017} extended this concept by simulating the outcome of the exogenous organic delivery to a WLP. One of the main advantages of WLPs compared to the subsea hydrothermal vent environment for abiogenesis is that the WLPs allow for wet-dry cycles, which promote polymerization and formation of oligonucleotide chains with lengths up to 300 nucleotides \citep{DaSilva2015}.
To drive this polymerization, all four essential RNA nucleotides or other monomers of the alternative RNA-predecessors listed above have to be present in sufficient concentrations to form the first oligomers.
\citet{Becker2018,Becker2019} showed in laboratory experiments that the formation of RNA nucleosides and nucleotides is plausible in WLPs during wet-dry cycles. Moreover, the polymerization of ribonucleotides was demonstrated in the presence of clays or salts \citep{Ferris1996,Ferris2004,DaSilva2015}, metal ion catalysts \citep{Orgel2004}, and in lipid bilayers \citep{Chakrabarti_ea94,Toppozini2013} in WLP settings (including geothermal fields and hot springs). 

The building blocks of the nucleotides and other vital prebiotic molecules have been exogenously delivered onto the Hadean and Eoarchean Earth during the late heavy bombardment, either by comets or asteroids \citep[see, e.g.,][]{Gomes2005}. Even nowadays, meteorites seed our planet with these organics. Theoretical models suggest that inward-drifting pebbles delivered water and carbon to the forming Earth from the outer nebular regions located beyond the orbit of the proto-Jupiter \citep{Johansen2021}. \citet{Kooten2021} developed a model based on the experimental studies of the CV (Vigarano-type) chondrite Leoville. They concluded that the forming Earth could have accreted a substantial fraction of the inward-drifting dust from the outer solar system, with a composition similar to CI (Ivuna-type) chondrites. Analysis of the ruthenium isotope \ce{^{100}Ru} in Eoarchean rocks and meteorites showed that the Earth's mantle is partly composed of the same material as carbonaceous chondrites \citep{Fischer-Godde2017,Fischer-Godde2020}.

Based on these recent findings, primitive carbonaceous chondrites could be promising candidates for seeding WLPs on the early Earth with necessary life building blocks, as compared to the carbon-rich interplanetary dust particles and comets \citep{Pearce2017,Dauphas2002}. Carbonaceous chondrites contain up to \SI{20}{\percent} water, and \SIrange{1}{5}{\percent} carbon by weight \citep{Mason1963}. The organic matter in the carbonaceous chondrites consists of an insoluble and a soluble part, with a ratio of roughly 70:30 \citep{Bitz1966}. The insoluble organic matter (IOM) consists of kerogen-like compounds made of many aromatic rings or chains. Most of the IOM is formed in the interstellar medium and protoplanetary disks \citep{Derenne2010,Alexander2017}. Still, laboratory experiments showed that IOM might also be formed in hydrothermically active carbonaceous chondrites alongside amino acid precursors \citep{Kebukawa2017}. It is even proposed that IOM could have played a role in the (pre-)biotic synthesis and abiogenesis, as it shows various similarities to melanin polymers in microorganism and could act as catalysts \citep{DIschia2021}. The soluble organic matter (SOM) consists of much smaller organic molecules important for the origin of life, such as purines and pyrimidines, carboxylic acids, amines, amides, alcohols, aldehydes, ketones, amino acids, etc.\ \citep[see, e.g.,][]{Gilmour2003,Pizzarello2006}. Carbonaceous chondrites also contain \ce{P}-rich minerals such as schreibersite, which can phosphorylate nucleosides, creating nucleotides \citep{Gull2015}.

The SOM also contains the nucleobases guanine, adenine, and uracil with measured concentrations of \SIrange{0.25}{515}{ppb} \citep[parts per billion;][]{vanderVelden1977,Stoks1979,Stoks1981,Shimoyama1990,Callahan2011,Pearce2015}. Recently, cytosine and thymine were also found with concentrations of \SIrange{1}{5}{ppb} \citep{Oba2022}. Furthermore, the sugar ribose, a fundamental building block of RNA, was recently identified, with concentrations of \SIrange{4.5}{25}{ppb} \citep{Furukawa2019}. 
These molecules could have been precursors of RNA-like polymers needed for abiogenesis.

In this study, we model abiotic reaction pathways and calculate the evolution and abundances of the nucleobases adenine (\textbf{A}), uracil (\textbf{U}, not to be confused with uranium), cytosine (\textbf{C}, not to be confused with carbon), guanine (\textbf{G}), and thymine (\textbf{T}) inside carbonaceous chondrite's parent bodies, following the previous study by \citet{Pearce2016}. We adopted the thermodynamic conditions inside various hypothetical parent bodies as computed in the study by K.\ Lange et al.\ (2022, in preparation), using a simplified version of their model here. This allowed us for the first time to determine from which parts of the parent bodies nucleobase-rich meteorites might have originated. The resulting nucleobase abundances were compared to the values measured in carbonaceous chondrites found on Earth.

In the next Section~\ref{sec:model}, we outline all reaction pathways considered in our model for the nucleobase synthesis in carbonaceous planetesimals. We outline how the Gibbs free energies of formation are used to perform the thermochemical equilibrium calculations for these nucleobases. Also, we present our new, more feasible approach to set up the initial concentrations of reactants, which better suits the formation history of these parent bodies. Further, we introduce the adopted planetesimal model by K.\ Lange et al.\ (2022, in preparation) and give a short summary of the nucleobase abundances found in carbonaceous chondrites. Next, in Section~\ref{sec:methods}, we describe our computational methods. In Section~\ref{sec:results}, we present and analyze the computed abundances and compare them to the previous results by \citet{Pearce2016} and measured values. In Section~\ref{sec:discussion}, we evaluate the necessity of our new approach to apply the appropriate initial concentrations of volatiles. We discuss how the overproduction of \textbf{U} in our model could be related to the low \textbf{C} and \textbf{T} abundances found in carbonaceous chondrites. Finally, conclusions about the relevance of our findings for the origin of carbonaceous chondrites with high nucleobase content and the origin of life follow in Section~\ref{sec:conclusions}.

\section{Model Setup}\label{sec:model}

To set up our model we first define the set of reaction pathways used to simulate the synthesis of the canonical nucleobases. This set was previously compiled by \citet{Pearce2015,Pearce2016} based on experimental studies and considerations of the environment in parent body planetesimals.

Second, we shortly review why chemical equilibrium is a suitable assumption in this planetesimal environment (see Section~\ref{sec:equilibrium}). We introduce the Gibbs free energies of formation, which allow us to run the thermodynamic calculations.

As we finally want to obtain predictions of the nucleobase abundances from the simulated reaction pathways, we need to understand the composition of the relevant reactants in the source material of carbonaceous planetesimals. We start from the measured composition of comets \citep[and references therein]{Mumma2011} as the most pristine and best-preserved reservoir of material in the outskirts of the solar system \citep{Rauer2008}. Guided by solar nebula models \citep{SW11,Drozdovskaya_ea16,Bergner_Ciesla21,Lichtenberg2021} and experimental insights from temperature-programmed desorption experiments \citep[J.~He (private communication)]{Ayotte2001,Collings2004,Bergner2022}, we derive initial conditions appropriate for the parent bodies of carbonaceous chondrites. This improvement upon the initial cometary abundances gives us a better representation of the amounts of reactants available to the chemical pathways in the carbonaceous planetesimals.

To run our calculations in a realistic environment, we introduce a full thermodynamic planetesimal model (K.\ Lange et al.\ 2022, in preparation).

Finally, we summarize the measured nucleobase abundances in actual carbonaceous chondrites \citep{Stoks1979,Stoks1981,Shimoyama1990,Callahan2011,Pearce2015,Oba2022}, which will serve as the benchmark to verify our simulated results.

\subsection{Reaction Pathways}\label{sec:reactions}

In the context of abiogenesis, all essential biomolecules have to form via abiotic reaction pathways, using only primordially occurring simple molecules as reactants.
The reaction pathways for the formation of the nucleobases \textbf{A} (\ce{C5H5N5}), \textbf{G} (\ce{C5H5N5O}), \textbf{C} (\ce{C4H5N3O}), \textbf{U} (\ce{C4H4N2O2}), and \textbf{T} (\ce{C5H6N2O2}) used in this study were taken from \citet[Table~1]{Pearce2016}.
In our model, we excluded reactions with formamide as a reactant since these reactions did not significantly contribute to the aqueous synthesis of the nucleobases \citep[nos.~24, 61, 49, and 63 in][]{Pearce2016}. 

\begin{deluxetable*}{lcll}
    \tablecaption{Potential reaction pathways within meteorite parent bodies.\label{tab:reactions}}
    \tablehead{
        \colhead{No.} & \colhead{Type\tablenotemark{a}} & \colhead{Reaction\tablenotemark{b}} & \colhead{Source(s)}} 
    \startdata
        \underline{\textbf{A}denine} & & & \\
        1 & FT & \ce{CO + H2 + NH3 ->[NiFe\mathrm{+||}Al2O3\mathrm{+||}SiO2] \textbf{A} + H2O} & \citet{Yang1971};\\
        & & & \citet{Hayatsu1968}\\
        3 & NC & \ce{5 HCN{}_{(aq)} -> \textbf{A}{}_{(aq)}} & \citet{Larowe2008}\\
        4 & NC & \ce{HCN + NH3 -> \textbf{A}} & \citet{Yamada1969};\\
        & & & \citet{Wakamatsu1966}\\
        6 & NC & \ce{5CO + 5NH3 -> \textbf{A} + 5H2O} & \citet{Hayatsu1968}\\
        7 & NC & \ce{HCN + H2O -> \textbf{A}} & \citet{Ferris1978}\\
        8 & NC & \ce{HCN + NH3 + H2O -> \textbf{A}} & \citet{Oro1961}\\
        \underline{\textbf{U}racil} & & & \\
        29 & NC & \ce{2HCN{}_{(aq)} + 2CH2O{}_{(aq)} -> \textbf{U}{}_{(aq)} + H2{}_{(aq)}} & \citet{Larowe2008}\\
        32 & NC & \ce{\textbf{C} + H2O -> \textbf{U} + NH3} & \citet{Robertson1995};\\
        & & & \citet{Garrett1972};\\
        & & & \citet{Ferris1968}\\
        \underline{\textbf{C}ytosine} & & & \\
        43 & FT & \ce{CO + H2 + NH3 ->[NiFe\mathrm{+||}Al2O3\mathrm{+||}SiO2] \textbf{C} + H2O} & \citet{Yang1971};\\
        & & & \citet{Hayatsu1968}\\
        44 & NC & \ce{3HCN{}_{(aq)} + CH2O{}_{(aq)} -> \textbf{C}{}_{(aq)}} & \citet{Larowe2008}\\
        \underline{\textbf{G}uanine} & & & \\
        51 & FT & \ce{CO + H2 + NH3 ->[NiFe\mathrm{+||}Al2O3\mathrm{+||}SiO2] \textbf{G} + H2O} & \citet{Yang1971};\\
        & & & \citet{Hayatsu1968}\\
        54 & NC & \ce{5HCN{}_{(aq)} + H2O -> \textbf{G}{}_{(aq)} + H2{}_{(aq)}} & \citet{Larowe2008}\\
        \underline{\textbf{T}hymine} & & & \\
        58 & NC & \ce{2HCN{}_{(aq)} + 3CH2O{}_{(aq)} -> \textbf{T}{}_{(aq)} + H2O} & \citet{Larowe2008}\\
        62 & NC & \ce{\textbf{U} + CH2O + HCOOH + H2O -> \textbf{T}} & \citet{Choughuley1977}\\
    \enddata
    \tablecomments{The pathways are taken from \citet[Table~1]{Pearce2016}, excluding reactions with formamide as reactant, since these reactions turned out to be negligible (nos.~24, 61, 49, and 63 there). All pathways were shown to produce the respective nucleobases in laboratory experiments under conditions present in parent body planetesimals, except nos.\ 3, 29, 44, 54, and 58, which are theoretical and thermodynamically favorable \citep{Larowe2008}. Sometimes real meteorite powder was used as a catalyst, e.g., \ce{NiFe} alloy powder \citep[\SI{90.2}{\percent} \ce{Fe}, \SI{7.1}{\percent} \ce{Ni}, \SI{0.46}{\percent} \ce{Co}, \SI{0.26}{\percent} \ce{P}, \SI{1}{\percent} \ce{C}, \SI{1}{\percent} \ce{S}, \SI{80}{ppm} \ce{Ga}, \SI{320}{ppm} \ce{Ge}, \SI{1.9}{ppm} \ce{Ir};][]{Buchwald1975} from the Canyon Diablo meteorite \citep{Hayatsu1968,Yang1971}.}
    \tablenotetext{a}{FT: Fischer-Tropsch, NC: Non-catalytic.} \tablenotetext{b}{$\mathrm{+||}$: One, or the other, or all catalysts used in laboratory experiments.}
\end{deluxetable*}

\vspace{-24pt}
All reactions in our model are summarized in Table~\ref{tab:reactions}. There are two types of processes, namely, Fischer-Tropsch (FT) and non-catalytic (NC) synthesis. FT reactions involve gaseous ammonia, carbon monoxide, and hydrogen and require a catalyst such as alumina or silica. NC reactions proceed in both gaseous and aqueous phases and do not require a catalyst.

This list of reactions was compiled in a previous study by \citet{Pearce2015} \citep[and extended by three pathways by][]{Pearce2016} mostly based on laboratory experiments, in which it was shown that each of these pathways forms the respective nucleobase in the presence of the corresponding reactants and catalysts (see references in last column of Table~\ref{tab:reactions}). The exceptions are the five pathways nos.~3,~29,~44,~54,~and~58 from \citet{Larowe2008}, which were found to be thermodynamically favorable in hydrothermal environments in theoretical calculations.

If the cited study proposed a chemical equation, we adopted the same equation in Table~\ref{tab:reactions}. On the other hand, if no balanced chemical equation was suggested, we follow the simple scheme
\begin{equation}
    \ce{reactants ->[catalysts] nucleobase + further products}.
\end{equation}

This explains why some of the reactions are not balanced, as the respective study did not follow a particular chemical equation, but only found the reactants to produce the nucleobase in their experiments. Later, this will become interesting when comparing the results from the different \textbf{A} pathways, as reaction no.~6 could be a balanced and NC version of no.~1, and no.~3 could be a balanced version of no.~4. We will also investigate the significance of \ce{NH3} in reaction nos.~4~and~8 compared to nos.~3~and~7, respectively. \citet{Pearce2016} already analyzed this aspect in their simulations, but we will re-examine their findings, as we changed the initial concentrations of reactants (see Section~\ref{sec:concs}) and will analyze if this affected the balance of the chemical equilibrium in the results.

Water forms alongside nucleobases in FT reactions \citep{Hayatsu1981}, and therefore, was added as a product to all FT reaction pathways.

\citet{Pearce2015,Pearce2016} reviewed a total of 63 reaction pathways as those applicable to the environment in planetesimals. Criteria for disregarding one of the pathways were the lack of necessary reactants, catalysts, or too high temperatures necessary for the pathways to proceed in the experiments. As meteorite parent body planetesimals might have incorporated the same material as present in comets \citep{Schulte2004,Alexander2011}, all pathways were ruled out if they required reactants or catalysts not found in comets. For example, pathway no.~43 \citep[in][]{Pearce2015} starting from cyanoacetaldehyde and urea \citep[e.g.,][]{Shapiro1999,Nelson2001} forming \textbf{C} was ruled out as cyanoacetaldehyde was not found in comets \citep[and references therein]{Mumma2011} or the interstellar medium \citep[and references therein]{Ballotta2021}. All five canonical nucleobases decompose in seconds at temperatures \makebox{$\gg\SI{400}{\kelvin}$} in aqueous conditions \citep{Levy1998}. Hence, pathways requiring these temperatures in the lab were ruled out. Thus, in this study, we use only experimentally verified pathways and disregarded other potential processes that have no laboratory data. In addition, we would like to keep our new model comparable to the \citet{Pearce2016} study with an emphasis on a more realistic asteroid model and more appropriate initial chemical composition of ices.

When running the chemical model, \textit{weak coupling} was assumed. This implies that different reactions do not compete for reactants and hence could be simulated independently. This is a safe assumption when reactants are simple molecules with much higher concentrations in the simulated environment than the synthesized complex reaction products \citep{Cobb2015}. In the previous studies by \citet{Cobb2015} and \citet{Pearce2016}, they found that the computed concentrations did not match the abundances measured in carbonaceous chondrites when multiple reactions were modeled simultaneously. Another reason why specific pathways may not have competed with each other for reactants is the porous structure of the carbonaceous chondrites, where different reaction processes could have been separated from each other by the thermochemical conditions (temperature, pressure, phase, presence of catalysts, etc.).

\citet{Pearce2016} attempted to simulate the synthesis of multiple nucleobases together by running several reaction pathways simultaneously. For example, when simulating the three FT reactions nos.~1,~43,~and~51 \citep{Hayatsu1968} together they found that only \textbf{C} was produced while the other nucleobases \textbf{A} or \textbf{G} stayed absent at any temperature in a \SIrange{0}{500}{\celsius} range. Their explanation for this was that the set of reactions in Table~\ref{tab:reactions} is the only experimentally verified one available, but might be incomplete and further molecules would be required, which were also detected as products in the laboratory experiments, e.g., urea, melamine, or guanidine \citep{Hayatsu1968}. This might lower the thermodynamic favorability of \textbf{C} and would allow for the other nucleobases to be produced. For further discussion regarding modeling competition between reactions, we refer to \citet[Section~4.4, Figure~8, and Appendix~B]{Pearce2016}.

To explore the possible transformation of nucleobases into each other, as a test case, reaction nos.~32~and~62 were also considered. As the resulting abundances of other pathways were used as initial concentrations for these two transformation reactions, this leads to a coupling of pathways. As this was not simulated simultaneously but individually, the resulting \textbf{U} and \textbf{T} abundances of nos.~32~and~62 represent only upper bounds, and the resulting \textbf{C} and \textbf{U} abundances resulting from other pathways might be lowered as they were consumed in these transformation reactions. This is discussed in Section~\ref{sec:overUlackCT}.

\subsection{Chemical Equilibrium in Planetesimals}\label{sec:equilibrium}

Our model assumes that all reactions in Table~\ref{tab:reactions} reach chemical equilibrium (see also Section~2.3 in \citet{Pearce2016} for an extensive discussion). This is also a prerequisite for the utilized chemical modeling software \textit{ChemApp} (see Section~\ref{sec:chemapp}). To achieve chemical equilibrium, the environment has to be in stable conditions over a duration that is longer than the time it needs to complete the reactions. The reactants involved in each reaction must be retained in the phase in which they react.

The planetesimal models (see Section~\ref{sec:planetesimal}) show that the interiors of typical parent bodies sustain temperatures allowing for liquid water and stay aqueous for several hundred thousand years in the smallest and earliest formed planetesimals, and up to several billion years in the biggest and latest formed ones. This was also confirmed by other studies. For example, planetesimals simulated in the study by \citet{Travis2005} with radii of \SIrange{50}{80}{\kilo\meter} and different initial content of \ce{^{26}Al} resulted in an aqueous phase longer than \SI{1}{\mega\year}. \citet{Lichtenberg2016} found for planetesimals with solid silicates and radii of around \SI{100}{\kilo\meter} similar timescales of several ten million years with a temperature range allowing for liquid water.

Compared to these aqueous timescales inside the planetesimals, the half-lives of the reactants involved and the duration of the reactions until their completion are orders of magnitude shorter. The half-life of \ce{HCN} in aqueous solution is less than ten thousand years \citep{Peltzer1984}. The \ce{HCN} measured in the Murchison meteorite was suggested not to be a free reactant, but was probably bound to \ce{-CN} salts formed in reactions with \ce{Fe^{2+}}, \ce{Mg^{2+}}, and \ce{Ca^{2+}} in the aqueuos phase of the planetesimal. Acidification of the meteorite samples was shown to lead to the release of this bound \ce{HCN} \citep{Pizzarello2012}. Consequently, all NC reactions involving \ce{HCN} in Table~\ref{tab:reactions} (nos.~3, 4, 7, 8, 29, 44, 54, and 58) used up all the freely available reactant and should have finished long before the planetesimals froze again.

The deamination of \textbf{C} to \textbf{U} in reaction no.~32 occurs on a timescale of ${\sim\SI{17000}{\year}}$ at \SI{0}{\celsius} and drops down in duration to \SI{3.5}{\hour} at \SI{165}{\celsius} \citep{Levy1998}. As these temperatures are easily reached in the planetesimal models and persist over longer timescales, this reaction should reach equilibrium. The recent detection of \textbf{C} in carbonaceous chondrites \citep{Oba2022} suggests that \textbf{C} is produced more rapidly via reactions nos.~43~and~44 than it is destroyed by deamination.

The reactions involving \ce{CO} and \ce{NH3} (nos.~1, 6, 43, and 51) formed the respective nucleobases in \SIrange{2}{288}{\hour} \citep{Hayatsu1968}, which is a negligible duration compared to the aqueous phase of the planetesimals. In the previous study \citep[Section~5.6]{Pearce2016} \ce{CO} was identified as a limiting reactant in these FT reactions and the lack of it in carbonaceous chondrites could verify that these reactions reached equilibrium.

Ultimately, reaction no.~62 transforming \textbf{U} into \textbf{T} with formaldehyde and formic acid finishes on the timescale of hours to days \citep{Choughuley1977}. Formaldehyde as the limiting reactant was found in carbonaceous chondrites \citep{Pizzarello2009,Monroe2011}, but it was suggested that this formaldehyde was not freely available, but tied up in other organic compounds or chemically adsorbed onto clays \citep{Pizzarello2009}, similar to \ce{HCN} as mentioned above.

At low concentrations, sufficient mixing of the dissolved reactants is required to achieve chemical equilibrium. The simulations of \citet{Travis2005} showed for planetesimals of \SIrange{50}{80}{\kilo\meter} strong hydrothermal convection inside these porous bodies. Up- and downwelling plumes took ${\sim\SI{50000}{\year}}$ to complete a full circulation, which would allow for many full cycles in the aqueous phase of our considered planetesimal models. Therefore, a well mixed and equilibrated distribution of the reactants can be assumed, even at the low concentrations of compounds in our model. When specific reactants necessary for a particular reaction pathway get available at a specific location inside the planetesimal due to convection, the short reaction times allow to restore chemical equilibrium quickly compared to the timescales of hydrothermal convection and period of aqueous conditions.

We conclude that chemical equilibrium is a well-constrained assumption in this context.

\subsection{Gibbs Free Energy of Formation}\label{sec:gibbs}

Every molecule has a Gibbs free energy of formation $\Delta G_f$ that varies with temperature and pressure. The lower the value of $\Delta G_f$, the higher the molecule's formation probability. For negative values of $\Delta G_f$, the molecule should form spontaneously (given all the necessary reactants are available).

$\Delta G_f$ as a function of temperature $T$ and pressure $p$ can be described by fitting the corresponding thermodynamical data (see Section~\ref{sec:gibbs_data}) with the function:
\begin{equation}\label{equ:gibbs_coeffs}
    \Delta G_f(T,p) = a + bT + cT\ln(T) + dT^2 + eT^3 + fT^{-1} + gp,
\end{equation}
where $a$--$g$ are the Gibbs coefficients \citep[see][Section~2.2]{Pearce2016}.

As the pressure dependence of $\Delta G_f$ is very marginal \citep{Cobb2015,Pearce2016}, the Gibbs coefficient $g$ can be neglected. For example, the relative percent difference of $\Delta G_f$ for \ce{HCN_{(aq)}} between \SI{1}{bar} and \SI{100}{bar} at \SI{300}{\kelvin} is \SI{0.4}{\percent} and similar for all the other molecules and temperatures considered. This makes the $\Delta G_f$ values pressure independent. The only pressure-dependent parameter of the model that one has to consider is the boiling point of water because the reactions are partly modeled in an aqueous phase. When water evaporates, we assume that the aqueous synthesis stops and the newly formed organic molecules are preserved at their current abundances. As the pressure does not influence the dynamics of the reactions, we did not calculate the lithostatic pressure as a function of radius inside the planetesimal. Instead, constant pressure of \SI{100}{bar} was assumed in the thermochemical calculations for the entire planetesimal. Only in the smallest considered planetesimals (see Section~\ref{sec:planetesimal}) with radii of only several kilometers is this pressure likely an overestimate. It could be that some reactions actually shut off as water starts to evaporate in the hottest central region (lower boiling point of water at lower pressures). This was not explicitly considered, and therefore, the calculated nucleobase abundances in the core regions of the smallest planetesimals are only valid if water stays in the liquid phase. However, when assuming that water stays liquid, the calculated abundances are correct as the Gibbs energies are nearly pressure-independent, as mentioned above. For the biggest considered planetesimals, the boiling point of water at \SI{100}{bar} was never reached.

The Gibbs coefficients $a$--$f$ are the input for the equilibrium chemistry software to model the chemical reactions (see Section~\ref{sec:chemapp}). To each reaction, one can assign a Gibbs free energy of the reaction $\Delta G_r$, which is defined as
\begin{equation}\label{equ:gibbs_react}
    \Delta G_r = \sum_{\mathrm{products}} \Delta G_f - \sum_{\mathrm{reactants}} \Delta G_f,
\end{equation}
and which has to be negative to be thermodynamically favorable. Otherwise, with a positive $\Delta G_r$, the reaction would require activation energy to proceed. This increases the system's total Gibbs free energy $\Delta G$, moving it away from the equilibrium (see below). The Gibbs free energy of the system $\Delta G$ is calculated by adding up each molecule's Gibbs free energy of formation,
\begin{equation}\label{equ:delta_gibbs}
    \Delta G = \sum_{\mathrm{all}} \Delta G_f.
\end{equation}

The chemical reactions will have essentially ceased when the system has reached equilibrium. At the concentrations of the reactants and products in the reaction at equilibrium, there can no longer occur a series of reactions that leads to a negative Gibbs free energy of the reaction $\Delta G_r$. We assumed chemical equilibrium and performed calculations to minimize the Gibbs free energy of the system $\Delta G$.

\subsection{Initial Concentrations of Reactants}\label{sec:concs}

In order to determine the initial concentrations of reactants, one faces the challenge that one does not have immediate access to the initial conditions present in carbonaceous chondrite parent bodies. As we always tried to feed as much experimental data and evidence as possible into our model, we started with the only preserved reservoir from the early stages of the solar system accessible to measurements, comets. However, comets formed in the outer solar system, and therefore, are not the representative reservoir at the correct radial location for the parent bodies of carbonaceous chondrites. Parent bodies of carbonaceous chondrites formed closer to the proto-Sun. The disk has a radial gradient in the concentrations of volatiles in ices as these start to desorb. To address this, we used a combination of solar nebula models \citep{SW11,Drozdovskaya_ea16,Bergner_Ciesla21,Lichtenberg2021} that predicted this volatile desorption, and accordingly, depletion in icy pebbles in the inner protoplanetary disk. We combine these models with insights from temperature-programmed desorption (TPD) experiments showing partial trapping of volatiles above their desorption temperatures.

\subsubsection{Comets}

Today, the most pristine and unmodified reservoir of material available are comets in our solar system \citep{Rauer2008,Mumma2011}. Meteorite parent body planetesimals might have incorporated some of the same material as present in comets \citep{Schulte2004,Alexander2011}. Therefore, comets represent a good starting point to obtain a prediction for the initial concentrations of reactants available in the source material of carbonaceous chondrites.

The remote studies of comets from Earth and \textit{in situ} measurements of 67P/Churyumov-Gerasimenko by the ROSINA mass spectrometer aboard the Rosetta mission showed that comets consist of dust and pristine chemical constituents that formed at very low temperatures \citep[${\lesssim\SI{30}{\kelvin}}$;][]{Mumma2011,2015A&A...583A...1L,2016SciA....2E0285A,2019MNRAS.490...50D,2019A&A...629A..84E}. Comets have formed in the outer region of the solar system beyond \SIrange{5}{20}{au}, including some source material originating from further inner regions \citep{Brownlee2006,Ciesla2012}, and were able to retain volatile species such as \ce{CO}, \ce{CH4}, and \ce{O2}. Even though they contain key prebiotic ingredients, including precursors of sugars and the amino acid glycine, it was estimated that comets might not have contributed much to the Earth's surface inventory of carbon, nitrogen, and water \citep{2016E&PSL.441...91M}. Instead, it is believed that carbonaceous chondrites have delivered a significant fraction of organics and volatiles to the Earth's upper crust and surface \citep{Dauphas2002,Bergin_ea15,Fischer-Godde2020,Kooten2021}.

\subsubsection{Solar Nebula Models}

Comets and asteroids represent two distinct pristine and partly processed matter reservoirs.
Unlike icy comets, most carbonaceous chondrites probably originate from parent bodies formed in the much warmer region of the solar nebula at ${\sim \text{\SIrange{2}{3}{au}}}$ \citep{1967GeCoA..31..747V,Lodders03,Martin2021}, as, e.g., the asteroid \mbox{19 Fortuna} was postulated to be the parent body source of CM (Mighei-type) meteorites \citep{Burbine2002}.
The formation and composition of the carbonaceous asteroids have been shaped by a complex history of the growth and fragmentation of solids starting from dust grains, the inward radial drift of pebbles, and the gravitational interactions with \mbox{(proto-)Jupiter} and Saturn at a later stage \citep{Gomes2005,2010Icar..207..744M,2011Natur.475..206W,2014prpl.conf..547J,2020A&A...637A...5E}.
According to current models, these \SI{100}{\km}-sized planetesimals could have been formed via streaming instability by the gravitational collapse of pebble-sized solids within several orbital timescales \citep{Johansen_ea07,2014prpl.conf..547J,Ormel2010,Klahr_Schreiber20}. 
Recently, this theoretical idea has been supported by the analysis of the measured chondral sizes in several CO (Ornans-type) chondrites \citep{Pinto2021}.  

The inner \SIrange{2}{3}{au} region in the solar nebula has been located right outside the water snowline at ${T \lesssim \SI{150}{\kelvin}}$, and well inside the evaporation zones of more volatile ices such as \ce{N2}, \ce{CO}, formaldehyde, and \ce{CH4} \citep{1977E&PSL..36....1W,2001Sci...293...64A,2001M&PS...36..671C,Lodders03,2005PNAS..10213755B,2018GeCoA.239...17B,2020ApJ...897...82V,Lichtenberg2021,Oberg_Bergin21}. 
The inward-migrating icy pebbles, out of which carbonaceous asteroids could have been formed via spontaneous gravitational collapse, should have experienced a substantial loss of moderate to very volatile species, even prior to later thermal metamorphism inside the parent bodies \citep[see, e.g.,][]{Piso_ea15}. Carbonaceous chondrite matrices were found to be significantly depleted in volatile species \citep{2005PNAS..10213755B}.
Therefore, the idea to use a pristine cometary volatile composition as the composition of volatiles in the icy pebbles may lead to a substantial overestimate of the concentrations of the initial material for organic syntheses inside the carbonaceous chondrite parent bodies. This may explain why in the study by \citet{Pearce2016} some nucleobases were severely overproduced compared to the concentrations measured in carbonaceous chondrites unless the total volatile content per unit solid mass would be lowered artificially to substantially lower values. 

The building blocks of carbonaceous chondrites, \si{\milli\meter}-sized icy pebbles, were slowly radially migrating and gradually losing their most volatile content due to thermal evaporation and diffusion through porous water ice layers.
For example, \ce{CO}, formaldehyde, and \ce{HCN} should be depleted in the water ice of the pebbles, because their desorption temperatures (\makebox{${\sim \text{\SIrange{20}{70}{\kelvin}}}$}) are much lower than that of water ice (\makebox{${\sim \text{\SIrange{140}{160}{\kelvin}}}$}, depending on gas pressure). 
Accordingly, a severe depletion of the volatiles in the water ice mantles of dust grains in the warm inner solar nebula was predicted by various physico-chemical nebula models \citep[see, e.g.,][]{Visser_ea09,SW11,Drozdovskaya_ea16,Bergner_Ciesla21,Lichtenberg2021}.

For example, in the extended model of \citet{SW11} both the laminar and turbulent mixing chemical models of the solar nebula show that at a final moment in time of \SI{5}{\mega\year} the abundances of \ce{CO}, \ce{HCN}, and formaldehyde ices become depleted by about \numrange{5}{8} orders of magnitude at the radius of \SI{3}{au} compared to the pristine-like values at \SI{20}{au}. One has to keep in mind that in this study and the other solar nebula models, it was assumed that the volatile ices are able to leave the water ice matrix freely.

\subsubsection{Trapping of Volatiles}

In TPD experiments under ultra-high vacuum conditions, the major fraction of these volatile ices remains trapped in the water ice. Most of the trapped volatiles then desorb during the so-called ``molecular volcano'' desorption as the water ice matrix changes from the amorphous to the crystalline structure at \makebox{$T\approx{}$\SIrange{150}{160}{\kelvin}} \citep{May2012,May2013a,May2013b}. The remaining fraction co-desorbs with water at higher temperatures of ${T \gtrsim \SI{160}{\kelvin}}$ \citep{Ayotte2001,Collings2004,Cuppen_ea17,Potapov_McCoustra21,Bergner2022}. As our prediction for the temperature present at \SIrange{2}{3}{au} in the early solar system is ${T \approx \SI{160}{\kelvin}}$ (see Section~\ref{sec:surf_temp}), the remaining volatile content in the source material of carbonaceous chondrites should be the amount left after the volcano desorption took place on the icy pebbles. Therefore, the remaining volatile content is everything that only co-desorbs with water at higher temperatures.

However, this trapping could be explained by the short timescales (hours--days) employed in these experiments, while icy pebbles in the solar nebula could have been gradually losing their volatile content over thousands of years, becoming more volatile-poor compared to the TPD results \citep{Cuppen_ea17,Potapov_McCoustra21}. 
As the temperatures when trapping occurs are above the desorption temperatures of the volatiles (see above), molecules should be able to desorb from the walls of the water ice pores and move by diffusion.
Indeed, the relatively efficient diffusion rates for the volatile ices such as \ce{CO}, formaldehyde, and \ce{CO2} at ${T \gtrsim \SI{90}{\kelvin}}$ were measured experimentally or predicted via molecular dynamics calculations \citep[and references therein]{C5CP00558B}. Diffusion parameters of \ce{CO}, \ce{CO2}, \ce{H2}, \ce{D2}, \ce{O2}, \ce{N2}, \ce{CH4}, and \ce{Ar} moving along pores inside amorphous water ice were measured \citep{He2017,He2018}. By desorbing and adsorbing again on the walls of the water ice pore channels, the volatile molecules would slowly diffuse through the porous crystalline water ice in a random walk and finally be lost to space before the pebbles form the first planetesimals by accretion.

This diffusion process might lead to a final volatile content that is somewhere in between the amounts found trapped after the volcano desorption in the TPD experiments and the many orders of magnitude lower amounts predicted by the solar nebula models. Therefore, the results from the TPD experiments and the solar nebula models represent only upper and lower constraints on the actual volatile content present in the accreted source material of carbonaceous chondrites, respectively.

Accurate modeling of this complex diffusion process of volatiles leaving slowly the porous water ice is beyond the scope of the present study. We used these constraints to get a first-order estimate of the volatile content in our planetesimals and used it to model the chemical pathways.

\subsubsection{Predicting the Initial Reactant Concentrations}\label{sec:pred_concs}

\begin{table*}
    \caption{Depletion of the volatiles \ce{CO}, \ce{H2}, \ce{HCN}, and formaldehyde measured by TPD experiments at \SI{161.3}{\kelvin} (see Section~\ref{sec:surf_temp}) or predicted by solar nebula models at \SIrange{2}{3}{au} in comparison to cometary regions.\label{tab:depletion}}
    \setlength\tabcolsep{3pt}
    \begin{tabularx}{\textwidth}{llclclc}
        \hline
        \hline
        \multicolumn{1}{c}{\multirow{2}[2]{*}{Volatile}} & \multicolumn{1}{c}{\multirow{2}[2]{*}{Name}} & \multicolumn{5}{c}{Depletion Factor} \tabularnewline
        \cmidrule{3-7}
        & & \multicolumn{2}{c}{TPD} & \multicolumn{2}{c}{solar nebula models} & {mean OOM\tablenotemark{a}} \tabularnewline
        \hline
        \ce{CO} & carbon monoxide & $2.75\times10^{-2}$ & J.~He (priv.\ comm.)\tablenotemark{b} & $\hphantom{4.25\times{}}{\sim{}}10^{-8\hphantom{0}}$ & \citet{SW11}\tablenotemark{e} & $10^{-4.8}$ \tabularnewline
        & & $\hphantom{0.00}\sim10^{-2}$ & \citet{Collings2004}\tablenotemark{c} & & & $10^{-5\hphantom{.0}}$ \tabularnewline
        & & $\hphantom{0.00}\lesssim10^{-2}$ & \citet{Ayotte2001}\tablenotemark{c} & & & $10^{-5\hphantom{.0}}$ \tabularnewline
        & & & & $\hphantom{4.25\times{}}{<{}}10^{-11}$ & \citet{Lichtenberg2021}\tablenotemark{f} & \tabularnewline
        & & & & $\hphantom{4.25\times}\ll10^{-2\hphantom{0}}$ & \citet{Bergner_Ciesla21}\tablenotemark{g} & \tabularnewline
        \hline
        \ce{H2} (\ce{D2}) & hydrogen & $3.17\times10^{-4}$ & J.~He (priv.\ comm.)\tablenotemark{b} & \multicolumn{2}{c}{No data} & No data \tabularnewline 
        \hline
        \ce{HCN} & hydrogen cyanide & $6.6\hphantom{0}\times10^{-2}$ & \citet{Bergner2022}\tablenotemark{d} & $\hphantom{4.25\times{}}{\sim{}}10^{-5\hphantom{0}}$ & \citet{SW11}\tablenotemark{e} & $10^{-3.1}$ \tabularnewline
        & & & & $\hphantom{4.25\times{}}{\sim{}}0\hphantom{0^{-00}}$ & \citet{Bergner_Ciesla21}\tablenotemark{g} & No data \tabularnewline
        \hline
        \ce{CH2O} & formaldehyde & \multicolumn{2}{c}{\multirow{2}{*}{No data}} & $\leq4.25\times10^{-3\hphantom{0}}$ & \citet{Drozdovskaya_ea16} & \multicolumn{1}{c}{\multirow{2}{*}{No data}} \tabularnewline
        & & & & $\hphantom{4.25\times{}}{\sim{}}0\hphantom{0^{-00}}$ & \citet{Bergner_Ciesla21}\tablenotemark{g} & \tabularnewline
        \hline
    \end{tabularx}
    \tablenotetext{a}{OOM: Order of magnitude. To calculate the OOM for \ce{CO} and \ce{HCN},  the solar nebula model by \citet{SW11} was used.}
    \tablenotetext{b}{See also Figures~\ref{fig:TPD_CO}~and~\ref{fig:TPD_H2} in the Appendix. In the experiments, the isotopes deuterium (molecular, \ce{D2}) and \ce{^{13}CO} were used to avoid contamination by residual gases in the ultra-high vacuum chamber.}
    \tablenotetext{c}{The integrated amount of the volatiles left above \SI{161.3}{\kelvin} was estimated from the published TPD spectra.}
    \tablenotetext{d}{This value from the the published plot was verified by the corresponding author.}
    \tablenotetext{e}{Extra modeling data from \citet{SW11} that was not published.}
    \tablenotetext{f}{As the radial range considered in this solar nebula model did not reach into the inner regions at \SIrange{2}{3}{au}, the volatile abundance there was approximated by linear extrapolation from the data at \SI{20}{au}.}
    \tablenotetext{g}{This model did not reach into the inner regions at \SIrange{2}{3}{au} and the values given here correspond to the radial distance of \SI{20}{au}.}
\end{table*}

\begin{deluxetable*}{llDcD}
    \tablecaption{Initial concentrations of reactants used for simulating the reaction pathways in Table~\ref{tab:reactions}.\label{tab:init_concs}}
    \tablehead{
    \colhead{Molecule} & \colhead{Name} & \twocolhead{Cometary Concentration\tablenotemark{a}} & \colhead{Depletion Factor\tablenotemark{b}} & \twocolhead{Predicted Concentration\tablenotemark{c}} \\
    \colhead{$i$} & \colhead{} & \twocolhead{${[\mathrm{mol}_i\cdot{}\mathrm{mol}_{\ce{H2O}}^{-1}]}$} & & \twocolhead{${[\mathrm{mol}_i\cdot{}\mathrm{mol}_{\ce{H2O}}^{-1}]}$}
    }
    \decimals
    \startdata
        \ce{H2O} & water & 1 & None & 1 \\
        \ce{CO} & carbon monoxide & 1.75$\times10^{-1}$ & $10^{-5}$ & 1.75$\times10^{-6}$ \\
        \ce{H2} & hydrogen & 1.75$\times10^{-1}$ & $10^{-4}$ & 1.75$\times10^{-5}$ \\
        \ce{NH3} & ammonia & 7.0\phn$\times10^{-3}$ & None & 7.0\phn$\times10^{-3}$ \\
        \ce{HCN} & hydrogen cyanide & 2.5\phn$\times10^{-3}$ & $10^{-3}$ & 2.5\phn$\times10^{-6}$ \\
        \ce{HCOOH} & formic acid & 7.5\phn$\times10^{-4}$ & None & 7.5\phn$\times10^{-4}$ \\
        \ce{CH2O} & formaldehyde & 6.6\phn$\times10^{-4}$ & $10^{-3}$ & 6.6\phn$\times10^{-7}$ \\
    \enddata
    \tablecomments{All concentrations are normalized to water.}
    \tablenotetext{a}{Concentrations spectroscopically measured in comets \citep[and references therein]{Mumma2011} and used in the previous studies by \citet{Cobb2015} and \citet{Pearce2016}.}
    \tablenotetext{b}{The cometary concentrations are adjusted to be more in line with the carbonaceous chondrites' pristine composition. The cometary concentrations are multiplied by a depletion factor to give the predicted concentration values listed in the last column.}
    \tablenotetext{c}{Concentrations predicted for the source material forming carbonaceous chondrite parent bodies. The concentrations listed here are the ones used as input parameters in the thermochemical equilibrium calculations.}
\end{deluxetable*}

\vspace{-24pt}
Table~\ref{tab:depletion} gives an overview over the available studies and their results concerning the depletion of the volatiles \ce{CO}, \ce{H2}, \ce{HCN}, and formaldehyde, as predicted for the region where carbonaceous chondrite parent bodies have been formed. Results from both TPD experiments and solar nebula models are listed.

\paragraph{Experimental TPD Results}

TPD spectra for \ce{^{13}CO} and \ce{D2} were provided by J.~He (private communication). This allowed us to accurately determine the remaining volatile fraction for the temperature that might have prevailed in the formation region of carbonaceous chondrite parent bodies (see Figures~\ref{fig:TPD_CO}~and~\ref{fig:TPD_H2} in the Appendix). These isotopes were used to avoid contamination by residual gases in the ultra-high vacuum chamber. As \ce{H2} is more volatile than \ce{D2}, one would expect less trapping for \ce{H2} in comparison to \ce{D2}. We show in Section~\ref{sec:variable_volatile} that the initial \ce{H2} concentration has no impact on the resulting nucleobase abundances even for two orders of magnitude lower values. Therefore, it is a safe assumption to use the data for \ce{D2} as a proxy for the trapping of \ce{H2} in our study. \ce{^{13}CO} is less volatile than \ce{^{12}CO}, and one would expect more efficient trapping for \ce{^{13}CO} than for \ce{^{12}CO}. However, the mass and size difference between \ce{^{13}CO} and \ce{^{12}CO} is very small and the difference in trapping is negligible. In the conducted experiments, a mixture of volatile ice and water ice was deposited at \SI{10}{\kelvin} in a ratio of 1:10, which is close to the \ce{CO} and \ce{H2} composition in comets (\mbox{$\sim$1:6}, see Table~\ref{tab:init_concs}). The other studies, e.g., by \citet{Ayotte2001} and \citet{Collings2004}, used a lower volatile component in the deposited ice of \mbox{$\sim$1:20} and \mbox{$\sim$1:100}, respectively. To prepare the sample, J.~He (private communication) directly deposited the \ce{CO}-\ce{H2O} ice mixture onto the substrate before warming up, as has been done in \citet{Collings2004}. In contrast, \citet{Ayotte2001} first deposited the porous amorphous water ice and then the \ce{CO} ice on top of the ice matrix.

\citet{Bergner2022} performed TPD experiments with \ce{HCN}. The cometary abundance of \ce{HCN} in water ice is 1:400 (see Table~\ref{tab:init_concs}). \citet{Bergner2022} used a higher ratio of 1:20 due to detection constraints in the experiments. We refer in their study to the amount of \ce{HCN} trapped above \SI{160}{\kelvin} in 55 monolayers of water ice.

\paragraph{Results of Solar Nebula Models}

The considered solar nebula models predicted significantly lower volatile concentrations in the inner regions of the protoplanetary disk, as they did not consider trapping in water ice. We used the extended model by \citet{SW11} as we had direct access to all chemical data, such as the \ce{CO} and \ce{HCN} ice concentrations in the inner \mbox{$\leq\SI{5}{au}$} region of the solar nebula. Additionally, we used the published data of \citet{Bergner_Ciesla21} and \citet{Lichtenberg2021} and extrapolated the values from \SI{20}{au} to \mbox{$\leq\SI{5}{au}$}. The abundances of formaldehyde ice in the inner region at \SIrange{1}{10}{au} as well as outer cometary regions \mbox{$>\SI{30}{au}$} were adapted from the study by \citet{Drozdovskaya_ea16}. This study allowed us to derive an upper limit for the formaldehyde ice depletion factor of \mbox{$\leq\num{4.25e-3}$} (compare to Table~4 in their publication).

\paragraph{Volatile Depletion Factors}

To estimate the depletion of the volatile ices in the inner solar nebula as compared to cometary ices, we take the $\log_{10}$ of the depletion factors obtained in TPD experiments and the solar nebula models, and average them (see Table~\ref{tab:depletion}). Unlike in both TPD experiments and solar nebula models, these mean order of magnitude (OOM) values roughly account for trapping of the volatile ices and their slow diffusion out of the bulk water ice over long physical timescales associated with the growth and radial drift of the icy pebbles in the solar nebula. We regard the mean OOM as a reasonable first step in obtaining a more realistic composition of the source material of icy planetesimals in the inner solar system. We used the results from the solar nebula model by \citet{SW11} for \ce{CO} and \ce{HCN}, since the other considered models have not presented results for the inner \SIrange{2}{3}{au}.

Table~\ref{tab:init_concs} provides the cometary abundances \citep[and references therein]{Mumma2011} for all the reactants relevant to model the chemical pathways in Table~\ref{tab:reactions}. To derive the extra depletion factor to be applied to the cometary abundances, we used the mean OOM values obtained from TPD experiments and solar nebula models, whenever possible. In summary, we found that the initial cometary concentrations of \ce{HCN}, formaldehyde, \ce{H2}, and \ce{CO} from \citet{Cobb2015} and \citet{Pearce2016} should be reduced by factors of \num{e-3} (\ce{HCN}), \num{e-3} (\ce{CH2O}), \num{e-4} (\ce{H2}), and \num{e-5} (\ce{CO}), respectively (see Table~\ref{tab:init_concs}). These factors were chosen in this approximate manner to account for the large uncertainties and necessary assumptions involved. There are large discrepancies between and within the TPD and solar nebula studies (see Table~\ref{tab:depletion}). Thus, these factors are rather approximate, but may provide a more realistic representation of the initial volatile content of carbonaceous chondrites than cometary values. We see it as the most realistic attempt justifiable in the context of the broad uncertainties involved.

Another possible depletion process is the radiogenically powered outgassing of volatiles from planetesimals formed from the icy pebbles in the protoplanetary disk. The thermodynamic model of K.\ Lange et al.\ (2022, in preparation) that we use here (see Section~\ref{sec:planetesimal}) considers sublimation and deposition processes of the volatiles \ce{CO}, \ce{CH4}, \ce{CO2}, \ce{NH3}, and water in planetesimals. Except for water, they find significant outgassing of these volatiles. For low-temperature models designed for comets, they see a depletion of the volatiles in the core and a deposition of ice at outer shells. For higher temperatures, as considered here for asteroids, a main portion of the volatile content is outgassing and leaves the planetesimal. Only water stays evenly distributed over the radius of the planetesimals. \citet{Lichtenberg2021} saw the same outgassing effect for \ce{CO}, \ce{CO2}, and water, with a radially even distribution and only moderate depletion of water in the evolved planetesimals.

By combining available experimental information, theoretical predictions, and the above arguments, our estimates of the depletion factors for the volatile ices in carbonaceous chondrites (w.r.t. to the cometary values) might be considered reasonable.

\subsection{Planetesimal Model}\label{sec:planetesimal}

One of the most significant advancements in our study is the direct coupling of equilibrium nucleobase chemistry and a comprehensive 1D thermodynamic planetesimal model. \citet{Cobb2015} and \citet{Pearce2016} only considered the parameters for a single planetesimal parent body to discuss their results without using a comprehensive model. They did not couple the models directly. Their reaction calculations were carried out for the entire temperature range where water is expected to be liquid. Since the Gibbs energies of the molecules are pressure independent, as mentioned in Section~\ref{sec:gibbs}, their assumption of the pressure of \SI{100}{bar} for each modeled planetesimal body was kept in our simulations. Consequently, the interesting temperature range where water can become liquid at \SI{100}{bar} is \SI{273}{K} (\SI{0}{\degree C}) to \SI{584}{K} (\SI{311}{\degree C}).

In our simulations, we used a more advanced thermal model of a planetesimal developed by K.\ Lange et al.\ (2022, in preparation). This model simulated the thermal evolution of planetesimals from their formation in the solar system until today. The thermal history for each considered planetesimal as a function of the radius was used for the thermochemical reaction simulations with \textit{ChemApp} (see Section~\ref{sec:chemapp}). This allowed us to track the evolution of the nucleobase abundances with the planetesimal's thermal evolution over time.

It is important to note that the planetesimal models represent a simplified model adapted to parent bodies of carbonaceous chondrites compared to the more complex models of K.\ Lange et al.\ (2022, in preparation).

For the planetesimal simulation, the object's radius, its surface temperature corresponding to its distance to the Sun, the time of formation after calcium-aluminium-rich inclusions (CAI), and its porosity $\phi$ were adjustable parameters.
CAI are inclusions found in carbonaceous chondrites \citep[first evidence found in Allende chondrite]{Lee1977}, defining the age of the solar system as the oldest dated solids. In particular, the aluminium part with the isotope \ce{^{26}Al} is the main source of radiogenic heating inside planetesimals. Therefore, the time of formation after CAI defines the formation time of the planetesimal after the start of the formation of the solar system. This is important as this determines the amount of accreted radiogenic isotopes and hence the heat energy available inside the planetesimal.

The radioactive decay of the two short-lived isotopes \ce{^{26}Al} as the primary contributor and \ce{^{60}Fe} (${\text{half-lives} \approx \si{\mega\year}}$), and the four long-lived isotopes \ce{^{40}K}, \ce{^{232}Th}, \ce{^{235}U}, and \ce{^{238}U} (${\text{half-lives} > \si{\giga\year}}$) were considered as the heat source inside the simulated planetesimals. 
We had access to several models of planetesimals with radii of \SIrange{3}{150}{\kilo\meter} and times of formation after CAI of \SIrange{0.5}{3.5}{\mega\year}. For the porosity, ${\phi = \num{0.2}}$ was used in all the models as this corresponds to the typical value in carbonaceous chondrites \citep{Mason1963}. Recent studies show coinciding ranges and similar average values, e.g., \SIrange{10}{30}{\percent} \citep{Flynn1999} or \SI{17}{\percent} \citep{Macke2011}.

The equations of energy, gas, and ice conservation were implicitly discretized on a 1D grid and solved, including sublimation and deposition of volatiles and a source term accounting for radiogenic heating. Sublimation and deposition are several orders of magnitude faster than gas transport. Therefore, we assumed that the local gas pressure is at the local equilibrium vapor pressure at all times. Then, the porous body is in a quasi-steady-state, where the difference between the incoming and outgoing gas flux in a grid cell is immediately compensated by sublimation or deposition as long as the ice is locally present. We neglected the transport of latent heat for the simplicity of the model so that the equations for temperature diffusion and gas diffusion can be decoupled. The decoupled quasi-steady-state temperature diffusion and gas diffusion equations are both tridiagonal, allowing us to solve them separately with the \textit{scipy.linalg.solve\_banded()} routine, part of the \textit{scipy} \textit{Python3} packages \citep[\url{https://github.com/scipy/scipy},][]{2020SciPy-NMeth}.

\subsubsection{Surface Temperature}\label{sec:surf_temp}

As mentioned above, an essential parameter of the adopted planetesimal model was its surface temperature. When assuming a black body, the body's distance to the Sun can be used to determine its surface temperature. The typical distance of asteroids today that are believed to be parts of the same parent bodies as carbonaceous chondrites is between \SIrange{2}{3}{au}. For example, \mbox{19~Fortuna} was postulated to be the parent body source of CM meteorites \citep{Burbine2002}. Accordingly, we assumed \SI{2.5}{au} as the semi-major axis $a$ from the Sun for a parent planetesimal of typical carbonaceous chondrites. The unattenuated incident power $P_{\mathrm{in}}$ on the planetesimal at the time when the solar nebula had disappeared can be roughly calculated as:
\begin{equation}
    P_{\mathrm{in}} = \frac{(1-A) L_{\odot}' \pi r^2}{4 \pi a^2},
\end{equation}
with the planetesimal's albedo $A$, the solar luminosity at that time $L_{\odot}'$, and the planetesimal's radius $r$. Further assumptions were a spherical planetesimal on a circular orbit and an isotropic energy output of the Sun. Using the Stefan-Boltzmann law, the radiated power $P_{\mathrm{out}}$ from the surface of the planetesimal is
\begin{equation}
    P_{\mathrm{out}} = 4 \pi r^2 \epsilon \sigma T^4,
\end{equation}
where $\sigma$ is the Stefan-Boltzmann constant, $T$ is the surface temperature in Kelvin, and $\epsilon$ is the gray body emissivity of the planetesimal. Equating ${P_{\mathrm{in}} = P_{\mathrm{out}}}$ gives
\begin{equation}
    T = \left( \frac{(1-A)L_{\odot}'}{\epsilon \sigma 16 \pi a^2} \right)^\frac{1}{4}.
\end{equation}

As the early Sun was fainter than today, for the considered times of formation after CAI of \SIrange{0.5}{3.5}{\mega\year}, $L_{\odot}'$ was assumed to be approximately \SI{70}{\percent} of the present-day solar luminosity $L_{\odot}$ \citep{Bahcall2001}. Assuming a black body (${A = 0}$, ${\epsilon = 1}$) and that there was no large radial migration of the parent bodies in the young solar system \citep[neglecting, e.g., possible effects due to the ``Grand Tack'' hypothesis at early times, see, e.g.,][]{Masset2001,2011Natur.475..206W}, we obtained ${T = \SI{161.3}{\kelvin}}$.

\subsubsection{Simulated Conditions Inside Carbonaceous Planetesimals}

% Figures for figure set stored in 'fig01set.zip'
\figsetstart
\figsetnum{1}
\figsettitle{Temperature evolution inside different planetesimals.}

\figsetgrpstart
\figsetgrpnum{1.1}
\figsetgrptitle{3\,km, 0.5\,Myr}
\figsetplot{planetesimal_time_evolution_3km.pdf}
\figsetgrpnote{Temperature evolution inside planetesimal with a radius of \SI{3}{\kilo\meter}, formed at \SI{0.5}{\mega\year} after CAI, and a porosity $\phi$ of 0.2. Temperature curves are given for different distances to the center inside the body. Reproduced from a simplified version of the model by K.\ Lange et al.\ (2022, in preparation).}
\figsetgrpend

\figsetgrpstart
\figsetgrpnum{1.2}
\figsetgrptitle{4\,km, 1\,Myr}
\figsetplot{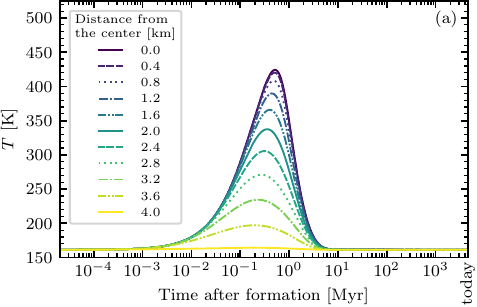}
\figsetgrpnote{Temperature evolution inside planetesimal with a radius of \SI{4}{\kilo\meter}, formed at \SI{1}{\mega\year} after CAI, and a porosity $\phi$ of 0.2. Temperature curves are given for different distances to the center inside the body. Reproduced from a simplified version of the model by K.\ Lange et al.\ (2022, in preparation).}
\figsetgrpend

\figsetgrpstart
\figsetgrpnum{1.7}
\figsetgrptitle{60\,km, 2.5\,Myr}
\figsetplot{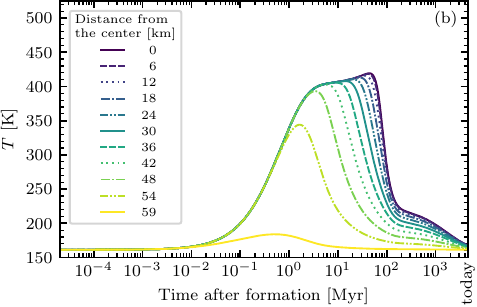}
\figsetgrpnote{Temperature evolution inside planetesimal with a radius of \SI{60}{\kilo\meter}, formed at \SI{2.5}{\mega\year} after CAI, and a porosity $\phi$ of 0.2. Temperature curves are given for different distances to the center inside the body. Reproduced from a simplified version of the model by K.\ Lange et al.\ (2022, in preparation).}
\figsetgrpend

\figsetgrpstart
\figsetgrpnum{1.8}
\figsetgrptitle{100\,km, 3.5\,Myr}
\figsetplot{planetesimal_time_evolution_100km.pdf}
\figsetgrpnote{Temperature evolution inside planetesimal with a radius of \SI{100}{\kilo\meter}, formed at \SI{3.5}{\mega\year} after CAI, and a porosity $\phi$ of 0.2. Temperature curves are given for different distances to the center inside the body. Reproduced from a simplified version of the model by K.\ Lange et al.\ (2022, in preparation).}
\figsetgrpend

\figsetgrpstart
\figsetgrpnum{1.9}
\figsetgrptitle{125\,km, 3.5\,Myr}
\figsetplot{planetesimal_time_evolution_124km.pdf}
\figsetgrpnote{Temperature evolution inside planetesimal with a radius of \SI{125}{\kilo\meter}, formed at \SI{3.5}{\mega\year} after CAI, and a porosity $\phi$ of 0.2. Temperature curves are given for different distances to the center inside the body. Reproduced from a simplified version of the model by K.\ Lange et al.\ (2022, in preparation).}
\figsetgrpend

\figsetgrpstart
\figsetgrpnum{1.10}
\figsetgrptitle{150\,km, 3.5\,Myr}
\figsetplot{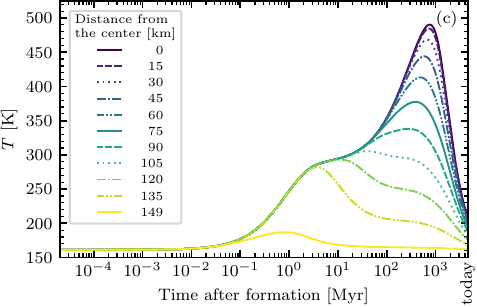}
\figsetgrpnote{Temperature evolution inside planetesimal with a radius of \SI{150}{\kilo\meter}, formed at \SI{3.5}{\mega\year} after CAI, and a porosity $\phi$ of 0.2. Temperature curves are given for different distances to the center inside the body. Reproduced from a simplified version of the model by K.\ Lange et al.\ (2022, in preparation).}
\figsetgrpend

\figsetend

% Example (sub-)figures (a), (b), (c) for Figure Set 1
\begin{figure}[p]
    \gridline{\fig{planetesimal_time_evolution_4km.pdf}{\columnwidth}{(a) ${\text{Radius} = \SI{4}{\kilo\meter}}$, time of formation after ${\text{CAI} = \SI{1}{\mega\year}}$.}}
    \gridline{\fig{planetesimal_time_evolution_60km.pdf}{\columnwidth}{(b) ${\text{Radius} = \SI{60}{\kilo\meter}}$, time of formation after ${\text{CAI} = \SI{2.5}{\mega\year}}$.}}
    \gridline{\fig{planetesimal_time_evolution_149km.pdf}{\columnwidth}{(c) ${\text{Radius} = \SI{150}{\kilo\meter}}$, time of formation after ${\text{CAI} = \SI{3.5}{\mega\year}}$.}}
    \figcaption{Temperature evolution inside different planetesimals with a porosity $\phi$ of 0.2 over time. Temperature curves are given for different radii inside the body (relative to the center). Reproduced from a simplified version of the model by K.\ Lange et al.\ (2022, in preparation). The complete figure set (10 images) is available (\href{https://doi.org/10.6084/m9.figshare.21545148}{figshare}, doi: \href{https://doi.org/10.6084/m9.figshare.21545148}{10.6084/m9.figshare.21545148}). It contains all planetesimal models with different radii and times of formation after CAI.\label{fig:planetesimals}}
\end{figure}

Figure~\ref{fig:planetesimals} shows the computed temperature evolution over time for a selection of the chondritic parent bodies with different sizes and times of formation after CAI. The supporting figure set (\href{https://doi.org/10.6084/m9.figshare.21545148}{figshare}, doi: \href{https://doi.org/10.6084/m9.figshare.21545148}{10.6084/m9.figshare.21545148}) shows all the available planetesimal models. The surface temperature derived above was used in the calculations and can be seen as the common baseline for all the temperature curves. 

In Figure~\ref{fig:planetesimals}a, most of the temperature curves inside a small, \SI{4}{\km}-sized planetesimal show a single strong temperature peak at ${t \sim \text{\SIrange{0.2}{0.6}{Myr}}}$ after formation, which was caused solely by the decay of the short-lived radionuclides. Due to the fast cooling of this small body, the influence of the long-lived isotopes was not noticeable.

However, in the larger bodies with a radius ${\gtrsim \SI{50}{\km}}$ the internal heat conduction was not efficient enough. Thus, their internal temperatures continued to increase after the half-life decay time of the short-lived isotopes (see Figures~\ref{fig:planetesimals}b~and~c). This lead to a second temperature bump at ${t \sim \SI{50}{Myr}}$ after formation in Figure~\ref{fig:planetesimals}b. Shortly thereafter, the temperature dropped due to the slow heat conduction and the declining abundances of the short-lived isotopes. Around \SI{200}{\mega\year} after the formation, thermal profiles became flatter since the long-lived isotopes still supplied heat (see Figure~\ref{fig:planetesimals}b). For an even larger planetesimal with a radius of \SI{150}{\km} in Figure~\ref{fig:planetesimals}c, the trapped heat from the short-lived isotopes started to overlap with the plateau caused by the long-lived ones in the core. This significantly extended the time span during which water could remain liquid, which is a key for the synthesis of the complex prebiotic molecules.

\subsubsection{Uncertainties of Planetesimal Model}\label{sec:discuss_planetesimal}

The utilized planetesimal model has several limitations. The latent heat of water ice was not considered in the planetesimal model. As the latent heat is significantly higher than the heat capacity of the bulk material, the phase transition from solid to liquid requires a considerable amount of energy. As a result, our planetesimal model overestimated the internal temperatures. In addition, crystallization of liquid water to amorphous ice releases heat during the cooling, resulting in less severe temperature decreases than computed by our model.

Furthermore, serpentinization reactions between rocks and water could produce additional heat and affect the calculated temperatures \citep[see, e.g.,][]{Gobi2017,Farkas-Takacs2021}. Moreover, if serpentinization happens, it could also distort the chemical reactions by binding the water necessary as the solvent for the chemistry or pushing the temperatures too high.

Sintering is another process that may occur. At high temperatures, a compactification of the rocks occurs, which allows even small porous bodies to reach high temperatures in a feedback loop. The increasingly compact body heats more easily, pushing sintering even more, while the material's melting point is still far from being reached. This strong feedback loop is expected to become important above \SI{700}{\kelvin} \citep{Henke2012,Henke2013}, which is well above the temperatures considered here. This was ensured by assuming the relatively late formation time after CAI (\SI{3.5}{\mega\year}) for the largest considered planetesimals of \SI{100}{\kilo\meter} radial size and larger. If these large planetesimals had formed earlier, they would reach temperatures above \SI{700}{\kelvin} and would probably experience strong sintering or even differentiation. No compactification and, therefore, change in porosity by sintering should be expected in the parent bodies considered here (thermal and aqueous metamorphosis could still change the porosity). Accordingly, fragments originating from any part of the bodies should have a chondritic structure and could be carbonaceous chondrites. At least the core region of early formed and large planetesimals is probably not of interest as the origin of fragments harboring prebiotic molecules, as these are only found in slightly aqueously altered meteorites (petrologic type $\sim 2$) that experienced no sintering. Sintering was only proposed as an explanation for the structure and petrologic type of some H- and L-chondrites \citep{Yomogida1983}, which probably descended from the core region of hotter (larger or earlier formed) parent bodies \citep{Henke2012} and were classified with high petrologic types ${>3.0}$.

The surface temperature used in the simulations was only roughly estimated, as explained in Section~\ref{sec:surf_temp}. The remaining dust in the protoplanetary disk could have partly shielded the light of the young Sun significantly. How much fainter the young Sun was is also uncertain. However, as the simulated reaction pathways were temperature-dependent only at high temperatures, a different surface temperature of the parent body only slightly shifts the minimum radius and maximum time of formation after CAI needed to start the synthesis of nucleobases. Hence, the resulting molecular abundances should not be affected much by such uncertainties.

The planetesimal model of \citet{Travis2005} showed hydrothermal convection in the interior of carbonaceous chondrite parent bodies. As our model did not consider this, convection could redistribute the water and, therefore, the reactants and products in our reaction pathways. In their study, they gave \SI{50000}{\year} as the rough timescale of a full convection cycle. As the aqueous phase of planetesimals lasts for several million years, in a first approximation, a well-mixed, roughly spherically and radially even water distribution could be assumed. Specifically for large \SI{100}{\kilo\meter}-sized and late formed planetesimals, heating by long-lived isotopes \citep[not considered by][]{Travis2005} extends the aqueous phase significantly. Nonetheless, considering convection could be interesting for a possible follow-up study.

\subsection{Meteoritic Abundances of the Canonical Nucleobases}

\citet{Pearce2015} reviewed the abundances measured in carbonaceous chondrites and gave ranges of \SIrange{1}{515}{ppb} for \textbf{G}, \SIrange{0.25}{267}{ppb} for \textbf{A}, and \SIrange{37}{73}{ppb} for \textbf{U}. Table~\ref{tab:meteorites} gives an overview over the nucleobase content found in several CM meteorites \citep[for abundances in carbonaceous chondrites of other types, see][]{Pearce2015}. The meteorite samples were ordered by decreasing \textbf{G} abundance.

\begin{deluxetable*}{llcDDD}
    \tablecaption{Abundances of \textbf{G}, \textbf{A}, and \textbf{U} in ppb measured in carbonaceous chondrites of type CM.\label{tab:meteorites}}
    \tablehead{
        \colhead{Meteorite} & \colhead{Sample Number} & \colhead{Type} & \twocolhead{\textbf{G}} & \twocolhead{\textbf{A}} & \twocolhead{\textbf{U}}
    }
        \decimals
        \startdata
        Murray & Murr. & CM2 & 515 & 236 & 37 \\
        Yamato & Yamato 74. & CM2 & 420 & \text{n.d.} & \text{n.d.} \\
        Lonewolf Nunataks & LON & CM2 & 244 & 30 & \text{n.d.} \\
        Murchison & ASU 1 & CM2 & 234 & 267 & 63 \\
        Yamato & Yamato 79. & CM2 & 230 & \text{n.d.} & \text{n.d.} \\
        Lewis Cliff & LEW & CM2 & 167 & 10 & \text{n.d.} \\
        Murchison & Smith. & CM2 & 56 & 5 & \text{n.d.} \\
        Meteorite Hills & MET & CM1 & 29 & 5 & \text{n.d.} \\
        Allan Hills & ALH & CM2 & 21 & 1 & \text{n.d.} \\
        Scott Glacier & SCO & CM1 & 2 & 4 & \text{n.d.} \\
        \enddata
        \tablecomments{n.d.: Not detected. The sample numbers given here are abbreviated and shortened versions of the ones given in the review by \citet{Pearce2015}. Please refer to there for the full sample numbers. Sources of measured abundances: \citet{Stoks1979,Stoks1981,Shimoyama1990,Callahan2011}.}
\end{deluxetable*}

Very recently, \citet{Oba2022} found the pyrimidines \textbf{C} and \textbf{T} in carbonaceous chondrites, with ranges of \SIrange{2}{5}{ppb} and \SIrange{1}{5}{ppb}, respectively. Nevertheless, it is important to note that they used a different technique to extract the SOM in comparison to the previous studies that were reviewed in \citet{Pearce2015}. \citet{Oba2022} used water for \SI{10}{\minute} under ultra-sonication at room temperature, whereas the previous studies, e.g., \citet{Callahan2011}, used \SI{95}{\percent} formic acid for \SI{24}{\hour} at \SI{100}{\celsius}. Using this more gentle extraction method, \citet{Oba2022} were able to preserve and detect the fragile \textbf{C} and \textbf{T} molecules for the first time, but measured systematically lower abundances for the other nucleobases \textbf{G}, \textbf{A}, and \textbf{U} in comparison to the previous studies. Therefore, the measured abundances are not comparable between the studies. Nevertheless, we will discuss these new findings in comparison to our simulated predictions.

\section{Computational Methods}\label{sec:methods}

To perform the thermodynamic calculations we used the chemical equilibrium software \textit{ChemApp} and as input the Gibbs energies from the \textit{CHNOSZ} database. Taking into account the typical properties of carbonaceous chondrite parent bodies, we calculated the resulting nucleobase abundances.

\subsection{Program Structure}

With our model, we calculated the abundances of nucleobases using the reaction pathways from Table~\ref{tab:reactions} and typical conditions found inside the parent bodies of carbonaceous chondrites. A detailed planetesimal heating model was adopted (see Section~\ref{sec:planetesimal}). Our model consisted of programs written in the four programming languages \textit{Python3}, \textit{C++}, \textit{FORTRAN}, and \textit{R}, which were combined into a single executable script. The source code, excluding the proprietary \textit{ChemApp} library, and including the data of the planetesimal models, is openly available on Zenodo and as a Git repository \citep[\url{https://github.com/klauspaschek/prebiotic_synthesis_planetesimal},][]{klaus_paschek_2021_5774880}.

\subsubsection{Equilibrium Chemistry Software}\label{sec:chemapp}

The thermochemical equilibrium calculations were performed using the software \textit{ChemApp} distributed by GTT Technologies \citep[\url{https://gtt-technologies.de/software/chemapp/},][]{Petersen2007}. The \textit{ChemApp} subroutines, provided as binaries in the \textit{FORTRAN} programming language, were called in a program written in \textit{C++}.
The compiled \textit{C++} source code and the \textit{FORTRAN} binary files containing the \textit{ChemApp} library were linked together using the compiler \textit{g++}, which is part of the GNU Compiler Collection (GCC; \url{https://gcc.gnu.org/}).
To use this mixed \textit{C++}/\textit{FORTRAN} code inside a \textit{Python3} script, it was compiled into a \textit{Cython} module usable in \textit{Python3} with the \textit{pybind11} library \citep[\url{https://github.com/pybind/pybind11},][]{pybind11}.
A \textit{Makefile} handles the necessary compiler flags and file extensions for easy building of the software with \textit{g++}.

\textit{ChemApp} requires the Gibbs coefficients from Equation~\ref{equ:gibbs_coeffs} for each reactant and product, their initial abundances, and the temperature and pressure in the system. To compute the reactant and product abundances, including the targeted molecules for each reaction at equilibrium, \textit{ChemApp} first breaks down the initial molecular abundances of the reactants into their elemental abundances (here carbon, hydrogen, oxygen, and nitrogen). Then it builds the system back up into the combination of reactants and product abundances that provides the minimum value of $\Delta G$ \citep[Equation~\ref{equ:delta_gibbs};][]{Pearce2016}.

\subsubsection{Gibbs Energy Data}\label{sec:gibbs_data}

The Gibbs energy data was obtained from the \textit{CHNOSZ} thermodynamic database \citep[version 1.3.6 (2020-03-16), \url{https://www.chnosz.net},][]{Dick2019}, a package for the programming language \textit{R}.
From these Gibbs energy data, the Gibbs coefficients used by \textit{ChemApp} were derived by fitting Equation~\ref{equ:gibbs_coeffs}.
By using the \textit{Python3} module \textit{rpy2} (\url{https://github.com/rpy2/rpy2}), the access of the data was managed from \textit{Python3} as an embedded \textit{R} session. Fitting of Equation~\ref{equ:gibbs_coeffs} to the data to obtain the Gibbs coefficients was done with the \textit{scipy.optimize.curve\_fit()} function, part of the \textit{scipy} \textit{Python3} packages \citep[\url{https://github.com/scipy/scipy},][]{2020SciPy-NMeth}.

For a more detailed explanation about obtaining the Gibbs coefficients, we refer to \citet[Section 4.1]{Cobb2015} and \citet[Section~3.1]{Pearce2016}.

\subsubsection{Calculation of Abundances}\label{sec:calc_reactions}

The calculated molecular abundances were scaled to represent the absolute concentrations in the planetesimals for easy comparison with the values measured in carbonaceous chondrites. It was assumed that the synthesized molecules have been preserved in the icy form inside the planetesimals after they cooled at the end of the radioactive heating (see Section~\ref{sec:planetesimal}). Furthermore, it was assumed that the water ice completely filled the pores within the parent body, to remain consistent with the previous studies by \citet{Cobb2015} and \citet{Pearce2016} \citep[who referred to][]{Travis2005}. By taking the porosity, the densities of ice and the bulk planetesimal material, and the molecular masses into account, the correct concentrations $Z_i$ were calculated as follows:
\begin{equation}\label{equ:ppb}
    Z_i = Y_i \cdot \phi \frac{\rho_{\mathrm{ice}}}{\rho_{\mathrm{rock}}} = X_i \frac{M_i}{M_{\ce{H2O}}} \cdot \phi \frac{\rho_{\mathrm{ice}}}{\rho_{\mathrm{rock}}},
\end{equation}
with $X_i$ being the resulting concentration from \textit{ChemApp} for the molecule $i$ in units $[\mathrm{mol}_i\cdot{}\mathrm{mol}_{\ce{H2O}}^{-1}]$, $M_i$ being the molar mass of the molecule $i$, $M_{\ce{H2O}}$ being the molar mass of water, $Y_i$ being the absolute concentration, $\phi$ being the porosity, $\rho_{\mathrm{ice}}$ being the density of ice, and $\rho_{\mathrm{rock}}$ being the density of the bulk planetesimal material.

\section{Results}\label{sec:results}

% Figures for Figure Set 2 stored in 'fig02set.zip'
\figsetstart
\figsetnum{2}
\figsettitle{Adenine abundances from simulations of the adenine reaction pathways.}

\figsetgrpstart
\figsetgrpnum{2.10}
\figsetgrptitle{150\,km, 3.5\,Myr}
\figsetplot{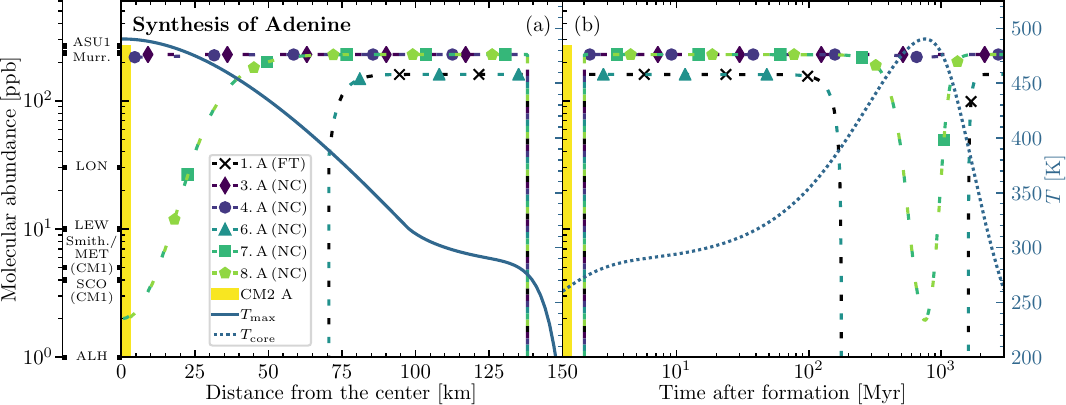}
\figsetgrpnote{Adenine abundances from simulations of the adenine reaction pathways nos.~1, 3, 4, 6, 7, and 8 in Table~\ref{tab:reactions}. Properties of planetesimal: ${\text{Radius} = \SI{150}{\kilo\meter}}$, densities ${\rho_{\mathrm{rock}} = \SI{3}{\gram\per\centi\meter\cubed}}$, ${\rho_{\mathrm{ice}} = \SI{0.917}{\gram\per\centi\meter\cubed}}$, porosity ${\phi = 0.2}$, and time of formation after ${\text{CAI} = \SI{3.5}{\mega\year}}$. All simulations were run at \SI{100}{bar}. In both panels (a) and (b) the left vertical axis corresponds to the abundances (dashed lines with symbols) and the right vertical axis corresponds to the temperatures from the planetesimal model (solid and dotted lines). In the figure legend, each reaction pathway no.\ is assigned to a symbol, the nucleobase produced is indicated with its initial letter in upper case, and the respective reaction type (FT or NC) in parenthesis. Each pathway is plotted as a dashed line with its assigned symbol. The left panel (a) shows the distribution of abundances for the maximum temperature $T_{\mathrm{max}}$ (solid line) reached at a specific distance from the center inside the planetesimal (center at the left and surface at the right). Adenine was synthesized at and below a distance of \SI{138}{\kilo\meter} from the center. The right panel (b) shows the temporal evolution of abundances at temperatures $T_{\mathrm{core}}$ (dotted line) in the center of the planetesimal (the same temperature evolution curve can be found in Figure~\ref{fig:planetesimals}c). Adenine synthesis started at \SI{2}{\mega\year} after formation. The shaded part of the abundance axis represents the range of adenine abundances measured in CM2 meteorites \citep{Callahan2011,Stoks1981}, and has no correlation to the radial location inside the object or the point in time (horizontal axes). On the left vertical abundance axis, the adenine abundances measured in individual meteorites are marked (using the abbreviated sample numbers in Table~\ref{tab:meteorites}). All meteorites are of type CM2 unless otherwise noted. Meteorites of type CM1 are indicated as such in parentheses below the abbreviated sample number.}
\figsetgrpend

\figsetend

% Example figure for Figure Set 2
\begin{figure*}[t]
    \includegraphics{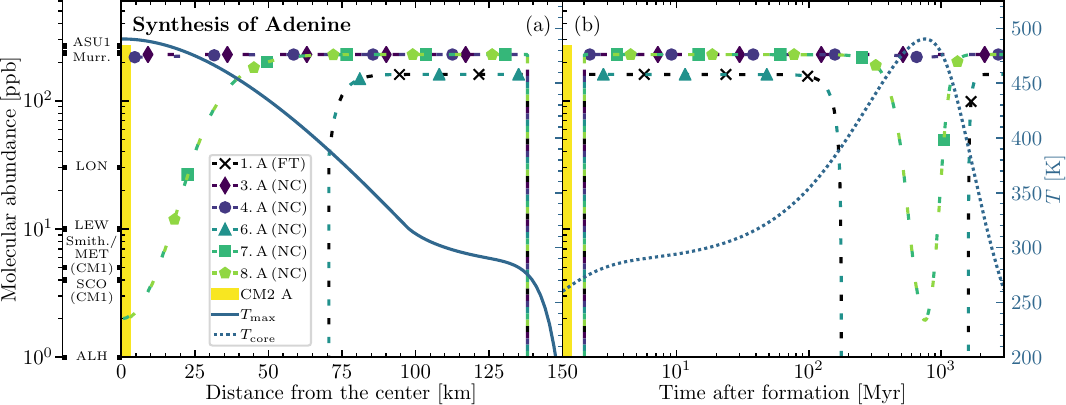}
    \figcaption{Adenine abundances from simulations of the adenine reaction pathways nos.~1, 3, 4, 6, 7, and 8 in Table~\ref{tab:reactions}. Properties of planetesimal: ${\text{Radius} = \SI{150}{\kilo\meter}}$, densities ${\rho_{\mathrm{rock}} = \SI{3}{\gram\per\centi\meter\cubed}}$, ${\rho_{\mathrm{ice}} = \SI{0.917}{\gram\per\centi\meter\cubed}}$, porosity ${\phi = 0.2}$, and time of formation after ${\text{CAI} = \SI{3.5}{\mega\year}}$. All simulations were run at \SI{100}{bar}. In both panels (a) and (b) the left vertical axis corresponds to the abundances (dashed lines with symbols) and the right vertical axis corresponds to the temperatures from the planetesimal model (solid and dotted lines). In the figure legend, each reaction pathway no.\ is assigned to a symbol, the nucleobase produced is indicated with its initial letter in upper case, and the respective reaction type (FT or NC) in parenthesis. Each pathway is plotted as a dashed line with its assigned symbol. The left panel (a) shows the distribution of abundances for the maximum temperature $T_{\mathrm{max}}$ (solid line) reached at a specific distance from the center inside the planetesimal (center at the left and surface at the right). Adenine was synthesized at and below a distance of \SI{138}{\kilo\meter} from the center. The right panel (b) shows the temporal evolution of abundances at temperatures $T_{\mathrm{core}}$ (dotted line) in the center of the planetesimal (the same temperature evolution curve can be found in Figure~\ref{fig:planetesimals}c). Adenine synthesis started at \SI{2}{\mega\year} after formation. The shaded part of the abundance axis represents the range of adenine abundances measured in CM2 meteorites \citep{Callahan2011,Stoks1981}, and has no correlation to the radial location inside the object or the point in time (horizontal axes). On the left vertical abundance axis, the adenine abundances measured in individual meteorites are marked (using the abbreviated sample numbers in Table~\ref{tab:meteorites}). All meteorites are of type CM2 unless otherwise noted. Meteorites of type CM1 are indicated as such in parentheses below the abbreviated sample number. The complete figure set (10 images) is available (\href{https://doi.org/10.6084/m9.figshare.21545148}{figshare}, doi: \href{https://doi.org/10.6084/m9.figshare.21545148}{10.6084/m9.figshare.21545148}). It contains the calculated abundances for the other available planetesimal models with different radii and times of formation after CAI.\label{fig:A}}
\end{figure*}

% Figures for Figure Set 3 stored in 'fig03set.zip'
\figsetstart
\figsetnum{3}
\figsettitle{Uracil and cytosine abundances from simulations of their reaction pathways.}

\figsetgrpstart
\figsetgrpnum{3.10}
\figsetgrptitle{150\,km, 3.5\,Myr}
\figsetplot{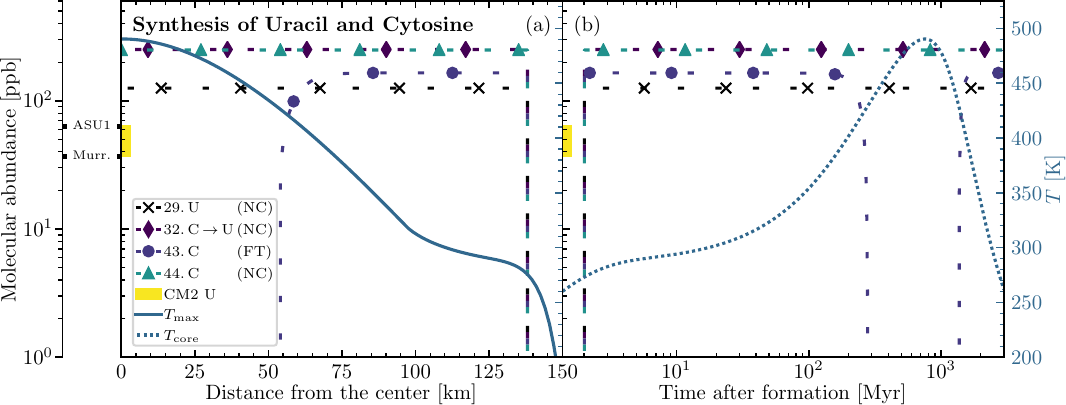}
\figsetgrpnote{Uracil and cytosine abundances from simulations of their reaction pathways nos.~29, 32, 43, and 44 in Table~\ref{tab:reactions}. Properties of planetesimal: ${\text{Radius} = \SI{150}{\kilo\meter}}$, densities ${\rho_{\mathrm{rock}} = \SI{3}{\gram\per\centi\meter\cubed}}$, ${\rho_{\mathrm{ice}} = \SI{0.917}{\gram\per\centi\meter\cubed}}$, porosity ${\phi = 0.2}$, and time of formation after ${\text{CAI} = \SI{3.5}{\mega\year}}$. All simulations were run at \SI{100}{bar}. In both panels (a) and (b) the left vertical axis corresponds to the abundances (dashed lines with symbols) and the right vertical axis corresponds to the temperatures from the planetesimal model (solid and dotted lines). In the figure legend, each reaction pathway no.\ is assigned to a symbol, the nucleobase produced (or transformation between nucleobases indicated by a reaction arrow) is indicated with its initial letter in upper case, and the respective reaction type (FT or NC) in parenthesis. Each pathway is plotted as a dashed line with its assigned symbol. The left panel (a) shows the distribution of abundances for the maximum temperature $T_{\mathrm{max}}$ (solid line) reached at a specific distance from the center inside the planetesimal (center at the left and surface at the right). Uracil and cytosine were synthesized at and below a distance of \SI{138}{\kilo\meter} from the center. The right panel (b) shows the temporal evolution of abundances at temperatures $T_{\mathrm{core}}$ (dotted line) in the center of the planetesimal (the same temperature evolution curve can be found in Figure~\ref{fig:planetesimals}c). Uracil and cytosine synthesis started at \SI{2}{\mega\year} after formation. The shaded part of the abundance axis represents the range of uracil abundances measured in CM2 meteorites \citep{Stoks1979}, and has no correlation to the radial location inside the object or the point in time (horizontal axes). On the left vertical abundance axis, the uracil abundances measured in individual meteorites of type CM2 are marked (using the abbreviated sample numbers in Table~\ref{tab:meteorites}).}
\figsetgrpend

\figsetend

% Example figure for Figure Set 3
\begin{figure*}[t]
    \includegraphics{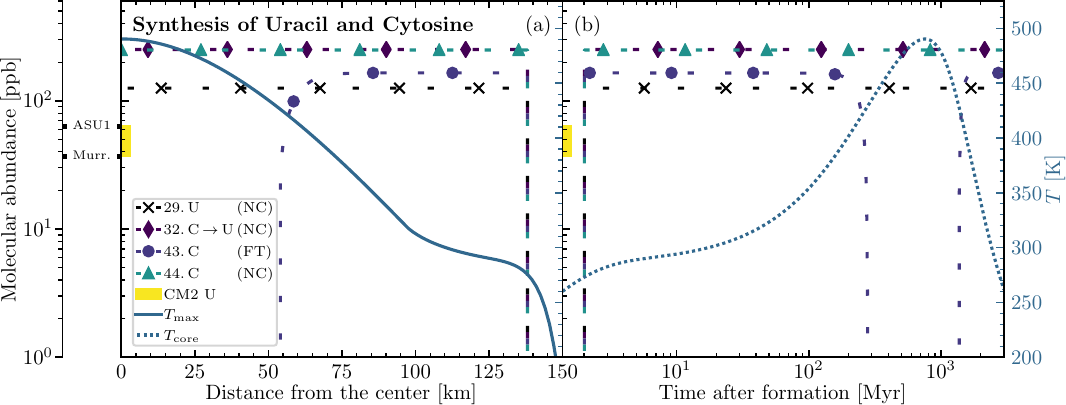}
    \figcaption{Uracil and cytosine abundances from simulations of their reaction pathways nos.~29, 32, 43, and 44 in Table~\ref{tab:reactions}. Properties of planetesimal: ${\text{Radius} = \SI{150}{\kilo\meter}}$, densities ${\rho_{\mathrm{rock}} = \SI{3}{\gram\per\centi\meter\cubed}}$, ${\rho_{\mathrm{ice}} = \SI{0.917}{\gram\per\centi\meter\cubed}}$, porosity ${\phi = 0.2}$, and time of formation after ${\text{CAI} = \SI{3.5}{\mega\year}}$. All simulations were run at \SI{100}{bar}. In both panels (a) and (b) the left vertical axis corresponds to the abundances (dashed lines with symbols) and the right vertical axis corresponds to the temperatures from the planetesimal model (solid and dotted lines). In the figure legend, each reaction pathway no.\ is assigned to a symbol, the nucleobase produced (or transformation between nucleobases indicated by a reaction arrow) is indicated with its initial letter in upper case, and the respective reaction type (FT or NC) in parenthesis. Each pathway is plotted as a dashed line with its assigned symbol. The left panel (a) shows the distribution of abundances for the maximum temperature $T_{\mathrm{max}}$ (solid line) reached at a specific distance from the center inside the planetesimal (center at the left and surface at the right). Uracil and cytosine were synthesized at and below a distance of \SI{138}{\kilo\meter} from the center. The right panel (b) shows the temporal evolution of abundances at temperatures $T_{\mathrm{core}}$ (dotted line) in the center of the planetesimal (the same temperature evolution curve can be found in Figure~\ref{fig:planetesimals}c). Uracil and cytosine synthesis started at \SI{2}{\mega\year} after formation. The shaded part of the abundance axis represents the range of uracil abundances measured in CM2 meteorites \citep{Stoks1979}, and has no correlation to the radial location inside the object or the point in time (horizontal axes). On the left vertical abundance axis, the uracil abundances measured in individual meteorites of type CM2 are marked (using the abbreviated sample numbers in Table~\ref{tab:meteorites}). The complete figure set (10 images) is available (\href{https://doi.org/10.6084/m9.figshare.21545148}{figshare}, doi: \href{https://doi.org/10.6084/m9.figshare.21545148}{10.6084/m9.figshare.21545148}). It contains the calculated abundances for the other available planetesimal models with different radii and times of formation after CAI.\label{fig:U/C}}
\end{figure*}

% Figures for Figure Set 4 stored in 'fig04set.zip'
\figsetstart
\figsetnum{4}
\figsettitle{Guanine and thymine abundances from simulations of their reaction pathways.}

\figsetgrpstart
\figsetgrpnum{4.10}
\figsetgrptitle{150\,km, 3.5\,Myr}
\figsetplot{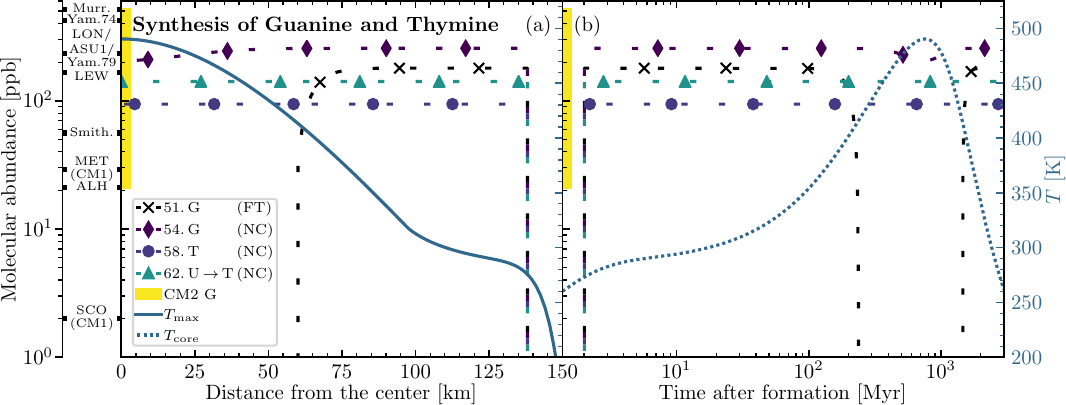}
\figsetgrpnote{Guanine and thymine abundances from simulations of their reaction pathways nos.~51, 54, 58, and 62 in Table~\ref{tab:reactions}. Properties of planetesimal: ${\text{Radius} = \SI{150}{\kilo\meter}}$, densities ${\rho_{\mathrm{rock}} = \SI{3}{\gram\per\centi\meter\cubed}}$, ${\rho_{\mathrm{ice}} = \SI{0.917}{\gram\per\centi\meter\cubed}}$, porosity ${\phi = 0.2}$, and time of formation after ${\text{CAI} = \SI{3.5}{\mega\year}}$. All simulations were run at \SI{100}{bar}. In both panels (a) and (b) the left vertical axis corresponds to the abundances (dashed lines with symbols) and the right vertical axis corresponds to the temperatures from the planetesimal model (solid and dotted lines). In the figure legend, each reaction pathway no.\ is assigned to a symbol, the nucleobase produced (or transformation between nucleobases indicated by a reaction arrow) is indicated with its initial letter in upper case, and the respective reaction type (FT or NC) in parenthesis. Each pathway is plotted as a dashed line with its assigned symbol. The left panel (a) shows the distribution of abundances for the maximum temperature $T_{\mathrm{max}}$ (solid line) reached at a specific distance from the center inside the planetesimal (center at the left and surface at the right). Guanine and thymine were synthesized at and below a distance of \SI{138}{\kilo\meter} from the center. The right panel (b) shows the temporal evolution of abundances at temperatures $T_{\mathrm{core}}$ (dotted line) in the center of the planetesimal (the same temperature evolution curve can be found in Figure~\ref{fig:planetesimals}c). Guanine and thymine synthesis started at \SI{2}{\mega\year} after formation. The shaded part of the abundance axis represents the range of guanine abundances measured in CM2 meteorites \citep{Callahan2011,Shimoyama1990,Stoks1981,vanderVelden1977}, and has no correlation to the radial location inside the object or the point in time (horizontal axes). On the left vertical abundance axis, the guanine abundances measured in individual meteorites are marked (using the abbreviated sample numbers in Table~\ref{tab:meteorites}). All meteorites are of type CM2 unless otherwise noted. Meteorites of type CM1 are indicated as such in parentheses below the abbreviated sample number.}
\figsetgrpend

\figsetend

% Example figure for Figure Set 4
\begin{figure*}[t]
    \includegraphics{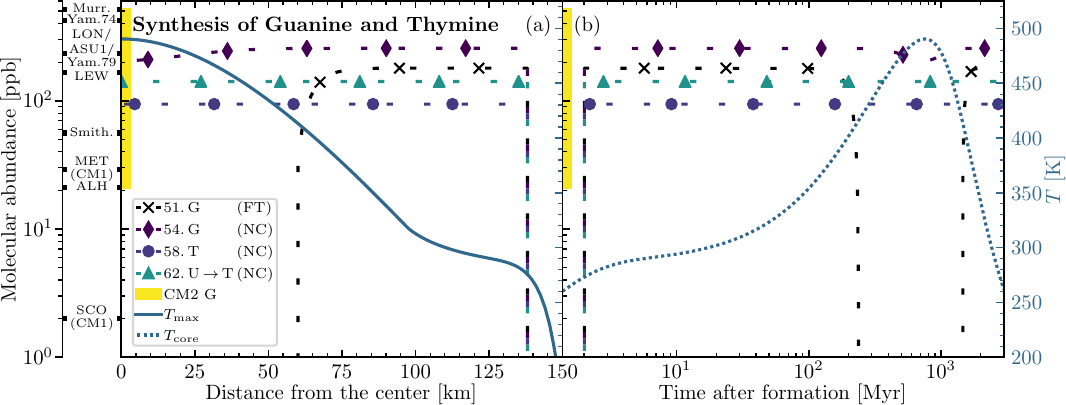}
    \figcaption{Guanine and thymine abundances from simulations of their reaction pathways nos.~51, 54, 58, and 62 in Table~\ref{tab:reactions}. Properties of planetesimal: ${\text{Radius} = \SI{150}{\kilo\meter}}$, densities ${\rho_{\mathrm{rock}} = \SI{3}{\gram\per\centi\meter\cubed}}$, ${\rho_{\mathrm{ice}} = \SI{0.917}{\gram\per\centi\meter\cubed}}$, porosity ${\phi = 0.2}$, and time of formation after ${\text{CAI} = \SI{3.5}{\mega\year}}$. All simulations were run at \SI{100}{bar}. In both panels (a) and (b) the left vertical axis corresponds to the abundances (dashed lines with symbols) and the right vertical axis corresponds to the temperatures from the planetesimal model (solid and dotted lines). In the figure legend, each reaction pathway no.\ is assigned to a symbol, the nucleobase produced (or transformation between nucleobases indicated by a reaction arrow) is indicated with its initial letter in upper case, and the respective reaction type (FT or NC) in parenthesis. Each pathway is plotted as a dashed line with its assigned symbol. The left panel (a) shows the distribution of abundances for the maximum temperature $T_{\mathrm{max}}$ (solid line) reached at a specific distance from the center inside the planetesimal (center at the left and surface at the right). Guanine and thymine were synthesized at and below a distance of \SI{138}{\kilo\meter} from the center. The right panel (b) shows the temporal evolution of abundances at temperatures $T_{\mathrm{core}}$ (dotted line) in the center of the planetesimal (the same temperature evolution curve can be found in Figure~\ref{fig:planetesimals}c). Guanine and thymine synthesis started at \SI{2}{\mega\year} after formation. The shaded part of the abundance axis represents the range of guanine abundances measured in CM2 meteorites \citep{Callahan2011,Shimoyama1990,Stoks1981,vanderVelden1977}, and has no correlation to the radial location inside the object or the point in time (horizontal axes). On the left vertical abundance axis, the guanine abundances measured in individual meteorites are marked (using the abbreviated sample numbers in Table~\ref{tab:meteorites}). All meteorites are of type CM2 unless otherwise noted. Meteorites of type CM1 are indicated as such in parentheses below the abbreviated sample number. The complete figure set (10 images) is available (\href{https://doi.org/10.6084/m9.figshare.21545148}{figshare}, doi: \href{https://doi.org/10.6084/m9.figshare.21545148}{10.6084/m9.figshare.21545148}). It contains the calculated abundances for the other available planetesimal models with different radii and times of formation after CAI.\label{fig:G/T}}
\end{figure*}

The abundances of products of the reaction pathways from Table~\ref{tab:reactions} were simulated using planetesimal models with different radii and times of formation after CAI. Figures~\ref{fig:A},~\ref{fig:U/C},~and~\ref{fig:G/T} show the adopted temperature curves (solid and dotted lines) in the largest considered planetesimal and the resulting target molecular abundances (dashed lines with symbols). The complete figure sets (3 $\times$ 10 images) are available (\href{https://doi.org/10.6084/m9.figshare.21545148}{figshare}, doi: \href{https://doi.org/10.6084/m9.figshare.21545148}{10.6084/m9.figshare.21545148}). The supporting figure sets show the molecular abundances calculated using the other available planetesimal models.
In each panel (a), for every distance from the center inside the planetesimal, the maximum temperature reached over the entire time evolution was used to calculate the molecular abundances. This corresponds to the conditions at which the synthesis of organics should be the fastest and hence the most efficient. 
The panels (b) show the time evolution in the center of the planetesimal, where the highest temperatures were reached. 

As can be clearly seen, nucleobase abundances for some reactions varied with temperature, especially at ${T \gtrsim \text{\SIrange{370}{450}{\kelvin}}}$ when several reactions shut off, as already found by \citet{Pearce2016}. At these high temperatures, the reactions nos.~1, 6, 7, 8, 32, 43, and 51 showed this behavior in their study. The same was observed in our study, except the synthesis in reaction no.~43 shut off already at lower temperatures (${\sim \SI{370}{\kelvin}}$) compared to \citet[${\sim \SI{530}{\kelvin}}$]{Pearce2016}. This was because of the lower initial concentrations of reactants used in our simulation, shifting the chemical equilibrium at these high temperatures compared to their previous study. Additionally, reaction no.~32 did not shut off at high temperatures, as explained below.

The model of \citet{Pearce2016} that used the cometary ice abundances of the reactants led to an overproduction of the nucleobase concentrations by many orders of magnitude in comparison to the measurements in carbonaceous chondrites (\SIrange{e5}{e6}{ppb} vs.\ \SIrange{0.25}{515}{ppb}, respectively). Only by scaling down the water content of the planetesimal by a factor of \num{5e-4}, the upper limits of the abundances found in the carbonaceous chondrites were reached in the simulations by \citet{Pearce2016}. 

Instead, in our model, the initial concentrations of the volatiles inside the parent bodies of carbonaceous chondrites were lowered by significant factors (but not the water abundance, see Table~\ref{tab:init_concs}). With these scaled-down input abundances, an overall satisfactory agreement between the simulated and measured values was achieved for \textbf{A} and \textbf{G} (see Figures~\ref{fig:A}~and~\ref{fig:G/T}). For \textbf{U}, the simulated abundances are higher, but by less than one order of magnitude (see Figure~\ref{fig:U/C}), and these values are close to an agreement now. This confirms the findings by \citet{Pearce2016} that the initial concentrations of reactants matter, in particular, those for \ce{HCN} and formaldehyde. We found that the concentrations of \ce{CO} and \ce{H2} are a limiting factor as well (see below).

The meteoritic \textbf{C} and \textbf{T} abundances of \SIrange{1}{5}{ppb} \citep{Oba2022} are lower than the amounts resulting from the individual pathways (see Figures~\ref{fig:U/C}~and~\ref{fig:G/T}). The recent discovery of \textbf{C} and \textbf{T} suggests that these nucleobases were either easily decomposed or were involved in further chemical reactions. Therefore, this discrepancy is to be expected, as pathways transforming these nucleobases (in combination with \textbf{U}) into each other were considered in our model (reactions nos.~32~and~62). Combining these pathways with the possible decomposition of \textbf{T} by hydrogen peroxide \citep[not considered here]{Shadyro2008} might explain the \textbf{C} and \textbf{T} abundances found in carbonaceous chondrites (see Section~\ref{sec:overUlackCT} for a detailed discussion).

\citet{Pearce2016} already investigated the importance of \ce{NH3} in reactions nos.~3~and~4 along with nos.~7~and~8. They found that \ce{NH3}, and hence, the respective pathways nos.~4~and~8 might not be important, since the resulting \textbf{A} abundances were identical. We re-examined these pathways to check if the change in initial concentrations shifted the chemical equilibrium. Figure~\ref{fig:A} shows that this was not the case and the resulting \textbf{A} abundances in the respective pairs of reactions stayed identical in comparison to each other. Therefore, we confirmed the findings of \citet{Pearce2016}. We further deduce from this that the proposed reaction pathway no.~3 in the theoretical study by \citet{Larowe2008} could explain which reaction is actually happening. It might be the balanced version of the reaction pathways no.~4 analyzed in the experimental work by \citet{Yamada1969} and \citet{Wakamatsu1966}, at least in the environment inside planetesimals. This means that \ce{NH3} either does not take part in the formation of \textbf{A} from \ce{HCN} or acts as a catalyst. This is consistent with the experimental findings for reactions nos.~7~and~8, which resulted in similar produced maximum \textbf{A} amounts with \citep[\SI{0.05}{\percent}, no.~8,][]{Oro1961} and without \citep[\SI{0.04}{\percent}, no.~7,][]{Ferris1978} \ce{NH3} as a reactant.

The FT synthesis reaction no.~1 of \textbf{A} and the NC reaction no.~6 start both from \ce{CO} and \ce{NH3} as reactants, but no.~1 also involves \ce{H2} and catalysts. Figure~\ref{fig:A} shows very similar \textbf{A} abundances for these two reactions, which might rule out the necessity of \ce{H2} in the formation of \textbf{A} in planetesimals. Nonetheless, \citet{Hayatsu1968} found that reaction no.~6 synthesizes \textbf{A} only at temperatures above \SI{500}{\celsius}, but \textbf{A} decomposes at these temperatures in less than \SI{1}{\second} \citep[Table~2]{Pearce2016}. The catalysts in reaction no.~1 might be necessary to form \text{A} at lower temperatures.

In summary, we found that the reaction pathways nos.~3~and~6 might be the balanced chemical equations for the unbalanced experimentally studied pathways nos.~4~and~1, respectively.

One significant new result from our study is the identification of the regions within the parent bodies from where nucleobase-containing meteorites have most likely been originated. Figure~\ref{fig:A}a shows that the reactions nos.~1~and~6 stopped producing \textbf{A} completely in the central regions of large planetesimals, and nos.~7~and~8 experienced a substantial decrease in the core region. If higher temperatures were reached inside the planetesimal (smaller time of formation after CAI or a bigger body), nos.~7~and~8 would stop completely. All other \textbf{A} reaction pathways would produce the same \textbf{A} abundances for the entire temperature range over which water remains liquid. However, when looking at Figure~\ref{fig:A}b, one can see that at later times, when the planetesimal cooled off, the shutdown reactions became again as productive as before the drop. This would suggest that these reaction pathways could have contributed to the production of the nucleobases in the core region. Still, one has to be careful as at such high temperatures, the formation of more refractory IOM could be favored, preventing the formation of \textbf{A} by irreversible ``trapping'' of the necessary reactants and by the improbable thermal destruction of IOM. Reaction nos.~43~and~51 showed the same behavior for \textbf{C} and \textbf{G} in Figures~\ref{fig:U/C}~and~\ref{fig:G/T}.

Further, \citet{Pearce2016} set the resulting abundance of \textbf{C} in reaction no.~43 as the input \textbf{C} concentration in reaction no.~32. They observed that \textbf{C} was converted nearly completely to \textbf{U}, which means that nearly all of the \textbf{C} in a planetesimal deaminated into \textbf{U}. They used the output \textbf{C} of reaction no.~43 as the input for reaction no.~32, as reaction no.~43 produced the highest amount of \textbf{C} in their study, in particular, higher than the ``direct'' synthesis of \textbf{U} from \ce{HCN} and formaldehyde in reaction no.~29. This situation has changed here, as we used lower initial concentrations of reactants. Here, reaction no.~44 (not no.~43) produced the highest amount of \textbf{C}. This was because the initial concentrations for \ce{HCN} and formaldehyde were lowered less than for \ce{CO} and \ce{H2} (see depletion factors in Table~\ref{tab:init_concs}), making a higher amount of reactants available to reaction no.~44 than to reaction no.~43. As mentioned above, the initial concentrations play a significant role and determine which reaction has the more abundant output. Therefore, we used the output \textbf{C} of reaction no.~44 as the input for the deamination reaction no.~32 to look for the upper limit of ``indirect'' formation of \textbf{U} via \textbf{C}, confirming the findings of \citet{Pearce2016} once again, as all \textbf{C} was deaminated to \textbf{U} (see nearly identical resulting abundances of \textbf{U} in reactions nos.~32 and 44 in Figure~\ref{fig:U/C}). This is a very important result for understanding the overproduction of \textbf{U} and the low amounts of \textbf{C} and \textbf{T} in carbonaceous chondrites (see Section~\ref{sec:overUlackCT}). Moreover, this explains why in our simulations, reaction no.~32 did not shut off at high temperatures as it did for \citet[${\sim \SI{530}{\kelvin}}$ there]{Pearce2016}, but remained productive up to the boiling point of water. This is the case because reaction no.~32 now directly followed reaction no.~44 since it took its output \textbf{C} as input. 

Taking a look at the individual meteorite samples in Table~\ref{tab:meteorites} and comparing them to the simulated nucleobase abundances allows us to determine from where inside the parent bodies a meteorite most likely originated. Further, the most suitable parent body for an individual meteorite can be inferred. To make this easier for the reader, the individual meteorites are marked on the abundance axis in Figures~\ref{fig:A},~\ref{fig:U/C},~and~\ref{fig:G/T} on the left.

Roughly three regimes can be identified inside the parent bodies when considering the temperatures and distances from the center at which particular types of chemical pathways shut off. The outermost and coolest regime ends when \ce{CO}-consuming (including FT) reactions (nos.~1,~6,~43,~and~51) shut off. This happens at \makebox{$\sim$\SIrange{350}{400}{\kelvin}}, corresponding to distances of \mbox{$\sim$\SIrange{55}{100}{\kilo\meter}} from the center in the \SI{150}{\kilo\meter}-sized model planetesimal. Second, at temperatures above \SI{400}{\kelvin} the water-consuming pathways no.~7,~8,~and~54 start to decrease or shut off (at \makebox{$\lesssim\SI{50}{\kilo\meter}$} distance from the center in the \SI{150}{\kilo\meter}-sized planetesimal). For \textbf{A}, one more regime can be identified. The high temperatures in the innermost core region allow only the remaining \ce{HCN}- and \ce{NH3}-consuming pathways no.~3~and~4 to be fully productive.

For example, the interplay between these regimes can be illustrated by comparing the meteorite fragments \textit{Murchison ASU 1} and \textit{Murray} (\textit{Murr.}). The \textbf{A} abundance of these meteorites is similar (\SI{267}{ppb} vs.\ \SI{236}{ppb}) and at the high end of abundances found in CM meteorites, but the \textbf{G} abundance is much lower in \textit{ASU 1} (\SI{234}{ppb} vs.\ \SI{515}{ppb}, see Table~\ref{tab:meteorites}). A possible explanation might be that these two meteorites originated from two different regions inside parent bodies. As \textit{Murray} is at the upper end of both \textbf{A} and \textbf{G} abundances this meteorite should have originated from the uppermost location where all pathways are active. This would correspond to distances of \makebox{$\gtrsim\SI{100}{\kilo\meter}$} up to \SI{138}{\kilo\meter} from the center in the \SI{150}{\kilo\meter}-sized planetesimal when taking the always frozen and inactive surface shell into account. Smaller planetesimals might also have been the parent body of \textit{Murray}, where it might have originated from the respective regime with temperatures \makebox{$<\SI{350}{\kelvin}$}. \textit{ASU 1}, in contrast, might have originated from the core of its parent body, which would explain the lower \textbf{G} abundance. In the \SI{150}{\kilo\meter}-sized planetesimal this corresponds to distances of \makebox{$\lesssim\SI{50}{\kilo\meter}$} from the center.

Other CM2 meteorite samples with rather low \textbf{A} and \textbf{G} abundances such as \textit{Lewis Cliff} (\textit{LEW}), \textit{Murchison Smith.}, and \textit{Allan Hills} (\textit{ALH}) might have been part of meteorites that originated from core regions as well. Another possibility might be that these meteorites originated from parent bodies with low initial \ce{CO} and \ce{H2} concentrations. As these were the most volatile reactants considered, the parent bodies of these meteorites might have formed inside the \ce{CO} snowline, leaving no \ce{CO} ice in the pebbles forming the parent body and shutting the respective \ce{CO}-consuming pathways down. Together with the other pathways, this might explain the lower \textbf{A} and \textbf{G} abundances found.

In general, an early-formed and small parent body is beneficial to explain these lower abundances in CM2 meteorites. These parent bodies allow for liquid water only over several hundred thousand years but are still able to reach high enough temperatures leading to shut-off pathways in their cores (see supporting figure sets of Figures~\ref{fig:A},~\ref{fig:U/C},~and~\ref{fig:G/T}; \href{https://doi.org/10.6084/m9.figshare.21545148}{figshare}, doi: \href{https://doi.org/10.6084/m9.figshare.21545148}{10.6084/m9.figshare.21545148}). This might explain why these meteorites of petrologic type 2 (only medium aqueous alteration of chondrules) might have experienced such high temperatures without accelerating aqueous alteration too much as the high temperatures only persisted for a short duration.

On the other hand, the CM1 meteorites \textit{Meteorite Hills} (\textit{MET}) and \textit{Scott Glacier} (\textit{SCO}) might have originated from core regions of larger and late-formed planetesimals as aqueous alteration might have been possible over long timescales (several million years), accelerated additionally by high temperatures in the core. This would explain why CM2 and CM1 meteorites with similar (low) \textbf{A} and \textbf{G} abundances were found.

Finally, we could explain the ranges of nucleobases found in CM2 and CM1 meteorites, respectively, with a scenario that all CM2 meteorites originated from the same early-formed and small parent bodies, and all CM1 from the same late-formed and large parent bodies. By being fragments from different distances from the center of the same two types of parent bodies, the whole range of nucleobase abundances can be explained.

\section{Discussion}\label{sec:discussion}

As the resulting amounts of nucleobases produced by various reaction pathways depend strongly on the initial concentrations of the volatiles, and these can only be estimated roughly, the presented results represent the best-case scenario. \citet{Pearce2016} used the measured abundances of the volatiles in comets, but this pristine matter may not represent the actual source material of the carbonaceous chondrites' parent bodies formed in a much warmer region of the solar nebula (as described in Section~\ref{sec:concs}). The application of solar nebula chemistry models and constraints from TPD experiments allowed us to constrain the abundances of volatiles in the regions of the carbonaceous chondrite parent bodies.

\subsection{Influence of Initial Volatile Concentrations}\label{sec:variable_volatile}

\begin{figure*}[p]
    \includegraphics{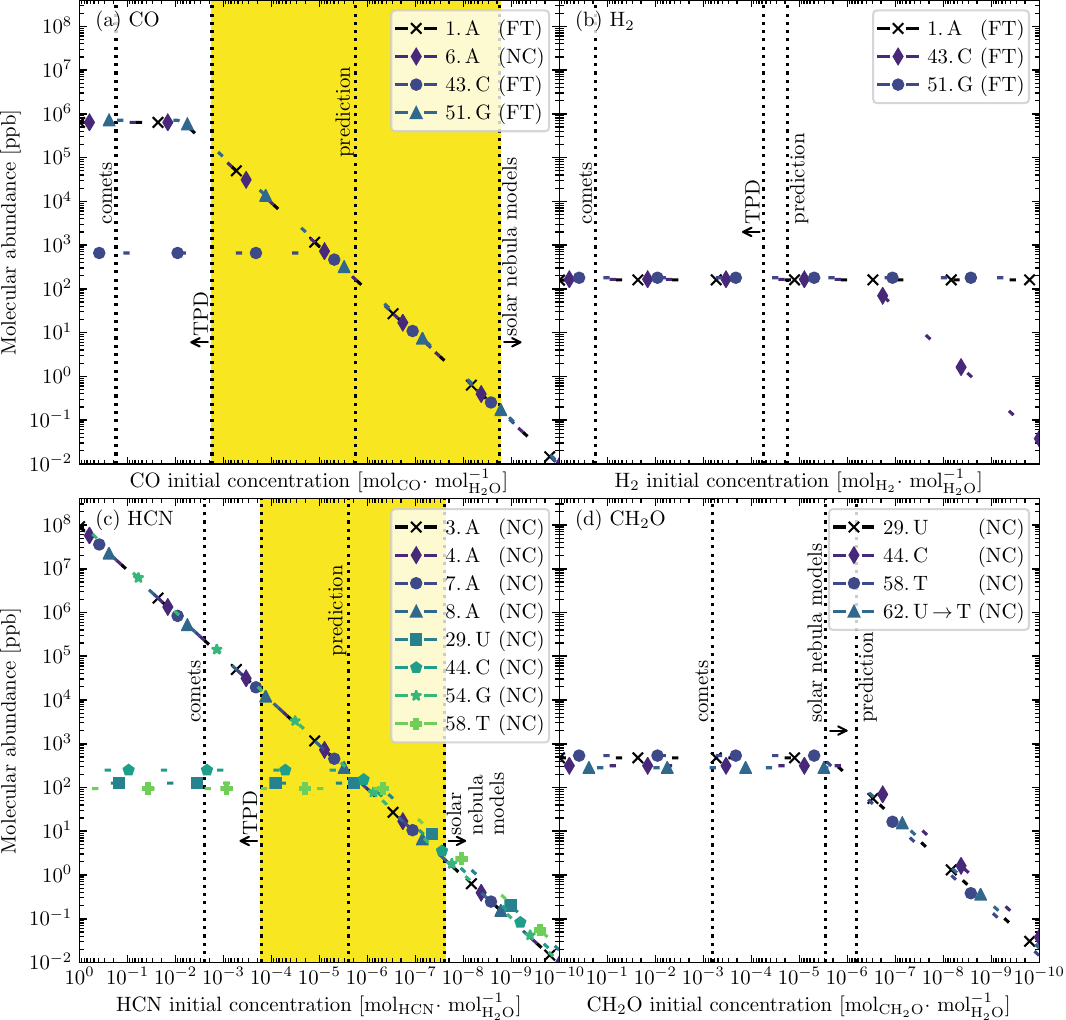}
    \figcaption{Nucleobase abundances from simulations of their reaction pathways in Table~\ref{tab:reactions} depending on variable initial concentrations of one of the volatile reactants \ce{CO}, \ce{H2}, \ce{HCN}, or \ce{CH2O}. All simulations were performed at \SI{0}{\celsius} and \SI{100}{bar}. Properties of planetesimal: Densities ${\rho_{\mathrm{rock}} = \SI{3}{\gram\per\centi\meter\cubed}}$, ${\rho_{\mathrm{ice}} = \SI{0.917}{\gram\per\centi\meter\cubed}}$, and porosity ${\phi = 0.2}$. Panel (a) shows how the resulting abundances (dashed lines with symbols) for the nucleobases in the pathways involving \ce{CO} depended on its initial concentration while the other involved reactants in each pathway were left at their predicted concentrations in Table~\ref{tab:init_concs}. Panels (b), (c), and (d) show the same for \ce{H2}, \ce{HCN}, and \ce{CH2O}, respectively. In the figure legends in each panel, each reaction pathway no.\ is assigned to a symbol, the nucleobase produced (or transformation between nucleobases indicated by a reaction arrow) is indicated with its initial letter in upper case, and the respective reaction type (FT or NC) in parenthesis. Each pathway is plotted as a dashed line with its assigned symbol. The vertical dotted lines indicate the initial volatile concentrations for the values i) measured in \textit{comets}, ii) reduced by the fraction found to be left at \SI{161.3}{\kelvin} in the measurements of \textit{TPD} experiments (if available), iii) the best \textit{prediction} obtained in the present study, and iv) (if available) obtained from \textit{solar nebula models} (see Tables~\ref{tab:depletion}~and~\ref{tab:init_concs}). If both TPD and solar nebula models were available for the respective volatile, the yellow shaded area indicates the range in between as the possible span of initial concentrations for the respective volatile in meteorite parent body planetesimals, and therefore also possible resulting nucleobase abundances.\label{fig:variable_volatile}}
\end{figure*}

The disagreement between experimental findings and theoretical models is visualized in Figure~\ref{fig:variable_volatile}. To explore the influence of the initial concentrations of the volatile reactants, the concentration of one of the volatiles was varied. The other involved reactants in each pathway were left at their predicted concentrations in Table~\ref{tab:init_concs}. Using standard conditions of \SI{0}{\celsius} and \SI{100}{bar}, the dependence of the resulting nucleobase abundances was calculated.

The amount of \ce{CO} and \ce{HCN} initially available in the planetesimals influenced the nucleobase abundances the most, whereas the depletion applied to the cometary abundances of \ce{H2} and formaldehyde had a minor effect. The chemical pathways show a linear dependence on the initial volatile concentration, and the majority of species reaches a saturation plateau for high volatile concentrations. A higher abundance of initial volatile reactants does not lead to more nucleobase production, in short, the nucleobase abundance does not depend on the initial volatile content any longer at high concentrations.

The reduction of the cometary \ce{H2} concentration did not change the results and would not even change them for two orders of magnitude stronger depletion (see Figure~\ref{fig:variable_volatile}b). Therefore, it is a safe assumption to use the TPD spectra of \ce{D2} in Figure~\ref{fig:TPD_H2} as a proxy for the initial \ce{H2} concentration in carbonaceous chondrite parent bodies (see Section~\ref{sec:pred_concs}). For formaldehyde, the depletion had the least impact, as the saturation plateau is close to the predicted concentration used (see Figure~\ref{fig:variable_volatile}d). For these two volatiles either results from TPD experiments or solar nebula models were taken into account (see Table~\ref{tab:depletion}), but as their influence on the nucleobase abundances is minor this imposes a negligible limitation on the validity of the presented results.

A change in the \ce{CO} and \ce{HCN} concentrations resulted in a one-to-one shift in the nucleobase abundances (see Figures~\ref{fig:variable_volatile}a~and~c). Clearly, the use of cometary concentrations provides an incorrect impression of the potential for nucleobase synthesis in parent bodies, and thus in carbonaceous chondrites, by several orders of magnitude. Nevertheless, the constraints on the volatile depletion by TPD experiments and chemical solar nebula models still allow for a broad spectrum of theoretically possible nucleobase abundances in carbonaceous chondrites (highlighted as yellow areas in Figure~\ref{fig:variable_volatile}). It is the task of future solar nebula models to include the experimental findings of volatile trapping and diffusion in a more appropriate way. The fact that the predicted abundances of \textbf{A} and \textbf{G} are in close agreement with the measured ones shows that considering all these effects is crucial to understanding why we find such abundances in meteorites.

\subsection{Overproduction of \textbf{U} and the Low Amounts of \textbf{C} and \textbf{T} in Carbonaceous Chondrites}\label{sec:overUlackCT}

A possible explanation for the slight overproduction of \textbf{U} is a combination of the reactions with nos.~32~and~62 and the fact that low \textbf{C} and \textbf{T} abundances have been found in carbonaceous chondrites \citep[\SIrange{1}{5}{ppb};][]{Oba2022}. The initial concentration of \textbf{C} in reaction no.~32 was set to the resulting one in reaction no.~44. Figure~\ref{fig:U/C} shows that the deamination reaction no.~32 was capable of converting \textbf{C} ultimately to \textbf{U}, as the resulting abundances in both reactions match each other perfectly for all temperatures. This could explain why little \textbf{C} has been found in carbonaceous chondrites, as it was probably formed, but was then almost completely transformed to \textbf{U} \citep[for another possible explanation for the low amount of \textbf{C} see][Section 5.5]{Pearce2016}. Furthermore, the resulting abundance of \textbf{U} in reaction no.~32 was used as the initial concentration in reaction no.~62, and half the amount of \textbf{U} was converted to \textbf{T}.
This could explain why the measured abundances of \textbf{U} in carbonaceous chondrites (shaded regions in Figure~\ref{fig:U/C}) are lower than the simulated values. This explanation may only work if there is an additional destruction pathway for \textbf{T} as low abundances were found in the meteorites, as mentioned above. \citet{Pearce2016} have proposed destruction of \textbf{T} by hydrogen peroxide \citep[\ce{H2O2};][]{Shadyro2008}.
Thus, when taking this possible explanation into account, our model is able to consistently explain the measured concentrations for all nucleobases.

However, because the reactions were handled separately, the coupling of reactions nos.~32~and~62 was not adequately simulated, and as a consequence, the resulting \textbf{U} and \textbf{T} abundances in those reactions merely reflect upper bounds. The trends for the \textbf{C} and \textbf{T} abundances in meteorites \citep{Oba2022} are qualitatively confirmed by these particular results, which only indicate trends with presumably lower actual abundances.

\subsection{Alternative Pathways, Isomers, Derivatives}

In addition to the canonical nucleobases studied in this paper, other purines and pyrimidines have been found in carbonaceous chondrites. \citet{Pearce2015} discussed the presence of xanthine, hypoxanthine, purine, {2,6-diaminopurine}, and {6,8-diaminopurine} in these meteorites, but due to the lack of experimentally verified pathways applicable to the environment of planetesimals, they were not considered further. If new experimental data become available it would of course be worth applying our model to these molecules. This could also shift the view on the canonical nucleobases, as these alternative molecules could be competing for reactants or be side products shifting the resulting abundances for the canonical nucleobases as presented here.

For the first time, \citet{Oba2022} found \textbf{C} and \textbf{T} in carbonaceous chondrites. Besides all the canonical nucleobases, they also found isomers of the canonical nucleobases and other derivatives, most prominently imidazole and several alkylated analogs. Most of these nucleobase and imidazole derivatives are missing in the \textit{CHNOSZ} Gibbs energy database and cannot be modeled yet.

The fundamental isomers of \textbf{G} and \textbf{C}, isoguanine and isocytosine, respectively, are critical as they might be direct side products in the formation process. On the other hand, the isomers have minor to negligible abundances in comparison to their canonical counterparts \citep{Oba2022}. The ratio of isoguanine to \textbf{G} was found to be \mbox{$\sim\SI{0.6}{\percent}$}, the ratio of isocytosine and \textbf{C} was \mbox{$\sim$\SIrange{10}{25}{\percent}} (depending on meteorite fragment). We did not consider these isomers in the present study because of their low abundances and the lack of experimental data.

We acknowledge this as a limitation of our study. When new experimentally verified pathways applicable in this context become available in the future, they should be added to the set of reactions. This might allow to determine if they have the potential to compete for the available reactants, and hence, also change the resulting canonical nucleobase abundances.

\subsubsection{Non-aqueous Formation Pathways and Inherited Organics}

We want to note that asteroids and meteorites might contain material that was formed in radiation-induced chemical reactions in the interstellar medium or the protosolar disk stage of the solar system and then inherited into these bodies. \citet{Ciesla2012} showed in their model that icy pebbles might have traveled to different parts of the protosolar disk. On their irregular paths, they were exposed to ultraviolet irradiation and thermal warming that, in laboratory studies, have been demonstrated to produce complex organics \citep[see, e.g.,][]{Nuevo2014,Materese2018,Oba2019}. This was also verified by theoretical quantum calculations \citep[see, e.g.,][]{Bera2010,Bera2016,Bera2017}. Analysis of the particles returned from comet 81P/Wild 2 by the Stardust mission showed that the nonvolatile part of the comet is made of an unequilibrated assortment of materials from both presolar and solar system origins. Some of these cometary particles were formed in the inner solar system \citep{Brownlee2006}. Therefore, the inheritance and mixing of radiation-induced organics into carbonaceous chondrite parent bodies might be imaginable. Still, we want to highlight the distinction between comets and carbonaceous chondrites. The amount of inherited organic material in carbonaceous chondrites is not precisely quantified, as they represent highly processes bodies in comparison to comets. Further, handling the environment for the synthesis of irradiation-induced organics, their incorporation into carbonaceous chondrite parent bodies, and their survival there requires an additional model completely different from the scenario investigated here. Calculating the synthesis and exact amount of possibly inherited organic material in carbonaceous chondrites is far beyond the scope of the present study.

\section{Conclusions}\label{sec:conclusions}

By lowering the initially high pristine cometary concentrations of volatile ices to values appropriate for the location of carbonaceous chondrite parent bodies and introducing a comprehensive planetesimal model of the parent bodies of carbonaceous chondrites, we successfully simulated the synthesis of all canonical nucleobases in meteorites in a consistent approach. Meteorites are time capsules from when the solar system formed ${\sim \SI{4.56}{\giga\year}}$ ago. They may reveal how the first building blocks of life could have been formed without the highly complex metabolism of the present life on Earth. The reaction pathways suggested by \citet{Pearce2015,Pearce2016} and listed in Table~\ref{tab:reactions} seem to explain reasonably well how nucleobases might have been formed abiotically.

The coupling of a thermochemical equilibrium model with a 1D thermodynamic planetesimal model allows to understand the history of the prebiotic synthesis in the carbonaceous chondrites' parent bodies over time. Further, it allows us to predict which regions of the parent bodies might have been the origin of the fragments containing the rich prebiotic molecule content that we find today in carbonaceous chondrites. The intermediate shell of a medium-sized parent body, which is not too close to the hot core and far enough away from the cold surface, should be the primary source of nucleobase-rich organics in carbonaceous chondrites. The fragments descending from this shell upon planetesimal collisions in the nebula might have been the source of the first prebiotic molecules on Earth, delivered to the young planet during the late heavy bombardment phase.

We confirm the finding of \citet{Pearce2016} that the ``indirect'' formation of \textbf{U} via \textbf{C} (deamination reaction no.~32) should have contributed more to the \textbf{U} abundance in carbonaceous chondrites than the ``direct'' synthesis from \ce{HCN} and formaldehyde in reaction no.~29, even for the lower initial concentrations of reactants used here. Reaction no.~44 and not reaction no.~43 (which shuts off at high temperatures) was the most effective one in producing \textbf{C} and therefore \textbf{U} via reaction no.~32. This implies that the hottest central parts of the parent bodies could have been the origin of \textbf{U}-rich carbonaceous chondrites (see Figure~\ref{fig:U/C}a). The reduction of the initial concentrations of reactants leads to a situation where the planetesimal cores contribute significantly to carbonaceous chondrites containing \textbf{U}.

This is in agreement with the \textbf{U} abundances measured in the meteorites \textit{ASU 1} and \textit{Murray} (see Table~\ref{tab:meteorites} and Figure~\ref{fig:U/C}). \textit{ASU 1} was found to originate from a location closer to the core of its parent body than \textit{Murray}, as inferred from the simulated abundances of \textbf{A} and \textbf{G} (see Section~\ref{sec:results}). This is reflected in the \textbf{U} abundances measured in these meteorites, as the core fragment \textit{ASU 1} contains a higher \textbf{U} concentration than \textit{Murray}. This confirms the above conclusion regarding the enhanced \textbf{U} synthesis in planetesimal cores due to reduced initial volatile concentrations. Thus, this validates the need to account for the depletion of volatiles and substantiates that our model can consistently identify the origins of individual meteorites.

Future studies of the solar nebula evolution with detailed gas-ice chemistry, growth, dynamics of the solids, gravitational collapse, and formation of asteroids from pebbles are needed to better constrain the original composition in organics in the carbonaceous chondrite parent bodies.

\citet{Pearce2016} found that the FT synthesis was the dominant mechanism to form nucleobases in planetesimals. This has changed here as the initial concentrations of \ce{CO} and \ce{H2} involved in FT synthesis were reduced more than for \ce{HCN} and formaldehyde involved in NC synthesis (see Table~\ref{tab:init_concs}), making now the NC synthesis the dominating mechanism for producing nucleobases in planetesimals. Still, FT and NC synthesis both contribute significantly.

An almost perfect match between the results of our calculations and the nucleobase concentrations measured in carbonaceous chondrites suggests that our new approach to adjusting the initial abundances of volatiles may indeed be correct \citep[see also][]{Dauphas2002}.

The sugar ribose, making up the backbone of RNA together with phosphates, was just found recently in carbonaceous chondrites \citep{Furukawa2019}. This molecule offers another excellent chance to model its formation in the parent bodies of the carbonaceous chondrites. In a follow-up study, we modeled its synthesis with our model, starting from formaldehyde and glycolaldehyde as reactants \citep{Paschek2022a}. Thermodynamic data and initial concentrations for glycolaldehyde are available from the previous study by \citet{Cobb2015}. As another key building block of the RNA world, ribose gives valuable insights into the possible exogenous origin of life on our planet and beyond.

\begin{acknowledgments}
K.P.\ acknowledges financial support by the Deutsche Forschungsgemeinschaft (DFG, German Research Foundation) under Germany's Excellence Strategy EXC 2181/1 - 390900948 (the Heidelberg STRUCTURES Excellence Cluster). K.P.\ is a fellow of the International Max Planck Research School for Astronomy and Cosmic Physics at the University of Heidelberg (IMPRS-HD). D.A.S.\ acknowledges financial support by the Deutsche Forschungsgemeinschaft through SPP 1833: ``Building a Habitable Earth'' (SE 1962/6-1). T.K.H.\ acknowledges financial support by the European Research Council under the Horizon 2020 Framework Program via the ERC Advanced Grant Origins 83 24 28. R.E.P.\ is supported by an NSERC Discovery Grant. The authors thank Cornelis P.\ Dullemond for his extensive help in understanding the planetesimal model and the background attached to its theory. We thank Jiao He for providing his measured data of TPD spectra to us, which helped to constrain the trapping and depletion of volatiles substantially. The authors would like to thank an anonymous reviewer for valuable comments and suggestions, which helped to improve the quality of the manuscript significantly. We would also like to thank Catharina Fairchild for her stylistic review of the manuscript.
\end{acknowledgments}

\newpage

\software{
    prebiotic\_synthesis\_planetesimal \citep[Klaus Paschek 2021, \url{https://github.com/klauspaschek/prebiotic_synthesis_planetesimal},][]{klaus_paschek_2021_5774880},
    ChemApp \citep[GTT Technologies 2020, \url{https://gtt-technologies.de/software/chemapp/},][]{Petersen2007},
    GNU Compiler Collection (GCC, Free Software Foundation, Inc.\ 2020, \url{https://gcc.gnu.org/}),
    pybind11 \citep[Wenzel Jakob 2016, \url{https://github.com/pybind/pybind11},][]{pybind11},
    CHNOSZ \citep[Jeffrey M.\ Dick 2020, version 1.3.6 (2020-03-16),  \url{https://www.chnosz.net},][]{Dick2019},
    rpy2 (Laurent Gautier 2008 - 2010, \url{https://github.com/rpy2/rpy2}),
    NumPy \citep[NumPy Developers 2005 - 2021, \url{https://numpy.org/},][]{harris2020array},
    SciPy \citep[Enthought, Inc.\ 2001 - 2002, SciPy Developers 2003 - 2019, \url{https://docs.scipy.org/doc/scipy/reference/},][]{2020SciPy-NMeth},
    Matplotlib \citep[John Hunter, Darren Dale, Eric Firing, Michael Droettboom and the Matplotlib development team 2002 - 2012, The Matplotlib development team 2012 - 2021, \url{https://matplotlib.org/},][]{matplotlib2007}
}

\clearpage
\appendix
\restartappendixnumbering

\section{TPD Spectra Provided with Permission by J.~He (Private Communication)}

\begin{figure*}[htp]
    \centering
    \includegraphics{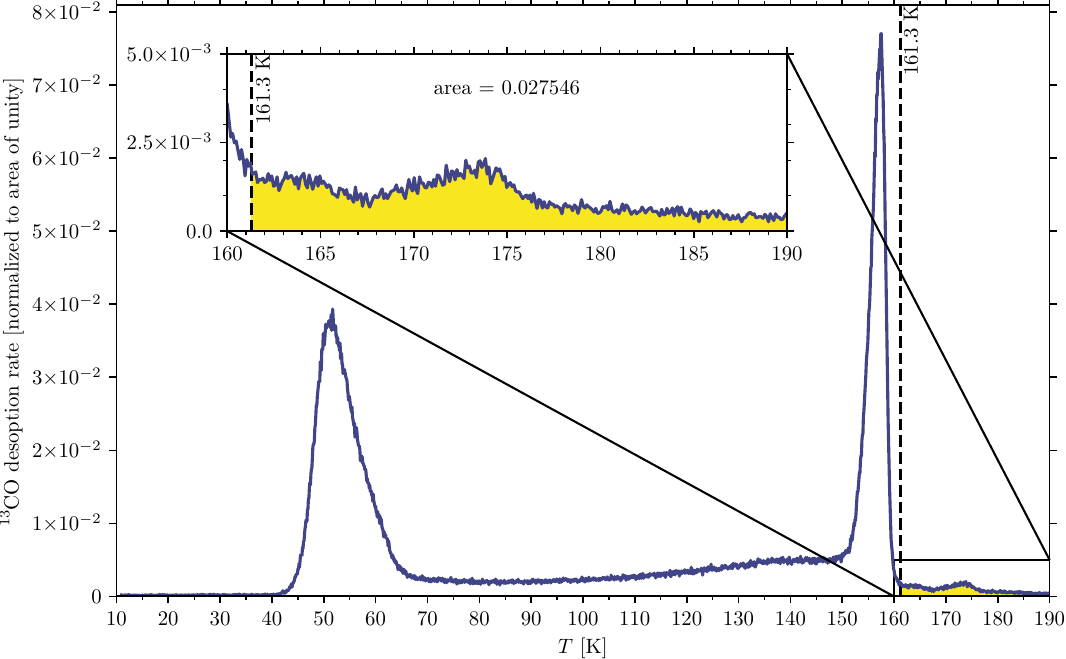}
    \caption{Results of TPD experiment for \ce{^{13}CO}-\ce{H2O} ice provided with permission by J.~He (private communication). The ratio between the volatile and water was 1:10, and the mixture was deposited at \SI{10}{\kelvin} and then warmed up while measuring the volatile desorption rate. The inset plot shows a zoom of the volatile trapped in the bulk water ice after ``molecular volcano'' desorption. The yellow shaded area indicates the integrated amount of volatile trapped above \SI{161.3}{\kelvin}, which corresponds to the temperature in the region of carbonaceous chondrite parent bodies at \SI{2.5}{au} in the solar nebula (see Section~\ref{sec:surf_temp}).}
    \label{fig:TPD_CO}
\end{figure*}

\begin{figure*}[hbp]
    \centering
    \includegraphics{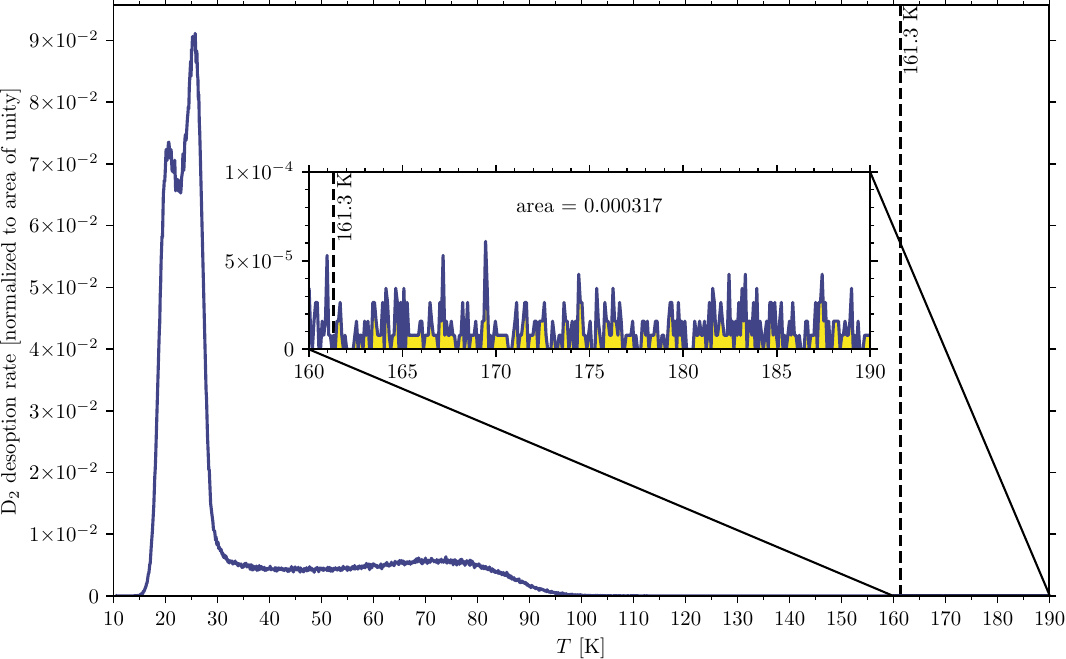}
    \caption{Results of TPD experiment for \ce{D2}-\ce{H2O} ice provided with permission by J.~He (private communication). The ratio between the volatile and water was 1:10, and the mixture was deposited at \SI{10}{\kelvin} and then warmed up while measuring the volatile desorption rate. In the zoomed-in inset plot, the yellow shaded area indicates the integrated amount of volatile trapped above \SI{161.3}{\kelvin}, which corresponds to the temperature in the region of carbonaceous chondrite parent bodies at \SI{2.5}{au} in the solar nebula (see Section~\ref{sec:surf_temp}).}
    \label{fig:TPD_H2}
\end{figure*}

\clearpage

\bibliography{main}
\bibliographystyle{aasjournal}



\end{document}